\documentclass[reprint,pre,notitlepage,superscriptaddress,amsmath,amssymb]{revtex4-2}

\usepackage{float}
\usepackage{graphicx}
\usepackage{rotating}
\usepackage[toc,page]{appendix}
\usepackage{dcolumn}
\usepackage{bm}
\usepackage{soul}
\usepackage[colorlinks=true,allcolors=blue]{hyperref}
\usepackage{color}
\usepackage{xcolor}
\newcommand{\p}[1]{\textbf{#1}}
\newcommand{\Lmu}{\mathcal{L}}
\newcommand{\Eqref}[1]{Eq.~\eqref{#1}}
\let\epsilon\varepsilon

\begin{document}

\title{Nucleation and time-reversal symmetry breaking in nonconserved scalar field theories}

\author{Noah Ziethen}
\email{niz20@cam.ac.uk}
\altaffiliation{These authors contributed equally to this work.}
\affiliation{DAMTP, Centre for Mathematical Sciences, University of Cambridge, Cambridge CB3 0WA, United Kingdom}
\author{Michalis Chatzittofi}
\email{mc2623@cam.ac.uk}
\altaffiliation{These authors contributed equally to this work.}
\affiliation{DAMTP, Centre for Mathematical Sciences, University of Cambridge, Cambridge CB3 0WA, United Kingdom}
\author{Michael Cates}
\affiliation{DAMTP, Centre for Mathematical Sciences, University of Cambridge, Cambridge CB3 0WA, United Kingdom}
\author{Cesare Nardini}
\affiliation{Service de Physique de l'\'Etat Condens\'e, CNRS UMR 3680, CEA-Saclay, 91191 Gif-sur-Yvette, France}
\affiliation{Sorbonne Universit\'e, CNRS, Laboratoire de Physique Th\'eorique de la Mati\'ere Condens\'ee, LPTMC, F-75005 Paris, France}

\date{\today}

\begin{abstract}
Classical nucleation theory (CNT) describes the formation of a stable phase from a metastable one in terms of a single reaction coordinate that corresponds to the radius of a nucleating droplet. In this work, we provide a full account of nonequilibrium nucleation theory (NNT), which generalizes CNT to non-equilibrium field theories with non-conserved order parameter. We present two equivalent derivations of the dynamics of the droplet radius: a stochastic route, based on a direct projection of the stochastic field equation onto the radial reaction coordinate, and a route based on the minimization of the Freidlin-Wentzell action. 
Crucially, the quasipotential barrier predicted by NNT differs from the one found when assuming the instanton to be the time-reversal of the relaxation dynamics. Whereas the interfacial density profile differs from that on the relaxation path, an analytical derivation of NNT remains possible using a careful definition of the reaction coordinate. This leverages the perturbative structure that (in common with CNT) emerges in the limit of large critical radius. 
We further derive with similar techniques the dynamics of capillary waves, whose stability is required for the CNT/NNT precept of a near-spherical droplet to prevail. After deriving our theory for generic non-conserved field-theories, we address two explicit examples: a non-equilibrium generalization of Model A (Active Model A), and a population dynamics model (with two choices of noise that each break time-reversal symmetry). In both cases, we validate our analytical NNT against numerical results obtained by action minimization, with excellent agreement. NNT provide a systematic framework for constructing nucleation theories in a broad class of non-equilibrium systems from active matter, reaction-diffusion systems and population dynamics.
\end{abstract}

\maketitle

\section{Introduction}

The spontaneous formation of a new phase within a metastable homogeneous state is a paradigmatic problem in statistical physics~\cite{Langer1967Jan}, arising across a wide range of physical, chemical, and biological systems~\cite{brayTheoryPhaseorderingKinetics, hohenbergTheoryDynamicCritical1977, RevModPhys.74.99, Cahn1958Feb, Allen1979Jun}. 
Classical examples arise in fluid--fluid demixing~\cite{onuki2002phase, brayTheoryPhaseorderingKinetics}, ferromagnetism~\cite{ishibashi1971note,rikvold1994metastable}, crystallization~\cite{vekilov2010nucleation}, phase-ordering systems~\cite{brayTheoryPhaseorderingKinetics,onuki2002phase}, reaction-diffusion systems~\cite{hinrichsen2000non,elgartRareEventStatistics2004, Halatek2018May,hellerPatternStabilityReactiondiffusion2026,cardy1998field}, and ecology~\cite{bastiaansenMultistabilityModelReal2018,Korniss2005Mar,DeAngelis2001Jan,Gandhi1999Sep,tanakaSpatialGeneDrives2017,giometto2021antagonism}. 

Nucleation phenomena also arise far from equilibrium, for instance in reaction--diffusion systems~\cite{elgartRareEventStatistics2004, Halatek2018May, Ouazan-Reboul2023Jul, schloglChemicalReactionModels1972}, ecological models with competing species~\cite{Korniss2005Mar, DeAngelis2001Jan, Gandhi1999Sep, bastiaansenMultistabilityModelReal2018, tanakaSpatialGeneDrives2017, giometto2021antagonism, cantrellSpatialEcologyReactionDiffusion2004}, subcellular condensates~\cite{lee2023size, shimobayashiNucleationLandscapeBiomolecular2021b}, and active matter~\cite{shankar2018defect,benvegnen2023metastability,catesClassicalNucleationTheory2023,lee2023size}.
In all these systems, transitions between metastable states are driven by fluctuations and become exponentially rare in the limit of small noise. Their probability $\mathcal{P}$ is governed by Large Deviation Theory (LDT), taking the form $\mathcal{P} \asymp \exp(-\mathcal{A}[\phi^*]/T)$, where $\asymp$ represents logarithmic equivalence, $\mathcal{A}$ is the rate function or action, $\phi^*$ is the most probable transition path (the instanton), and the `noise temperature' $T$ controls the strength of fluctuations~\cite{Freidlin1998, Touchette2009Jul, Bouchet2016Jun}. 

Analytical results on rare events in nonequilibrium systems are  scarce, for two reasons. First, the absence of a free energy blocks the calculation of transition probabilities by simply comparing the free energy of the metastable state to that of the lowest saddle connecting it to the stable one. Second, it is well-established that, unless detailed balance is respected, the instanton from the metastable state to the saddle is not the time-reversal of the relaxation path from the saddle to the metastable state~\cite{Touchette2009Jul,Freidlin1998,Bouchet2016Jun}. Sophisticated numerical techniques have therefore been developed to obtain the instantons and the transition probabilities~\cite{vanden2012rare,PhysRevX.13.041044, bucklew2004introduction, giardina2006direct, lecomte2007numerical, heymann2008, grafke2015instanton, grafke2019numerical, nemoto2014computation, ferre2018adaptive,  yan2022learning, PhysRevResearch.6.043110,Simonnet2023Oct,PhysRevLett.133.038301}.

Nonetheless, there is much insight to be gained from an analytic theory whenever it can be tractably developed.
The main paradigm for this, relevant in a large class of equilibrium systems, is Classical Nucleation Theory (CNT)~\cite{Langer1967Jan,oxtobyHomogeneousNucleationTheorya, DebenedettiPabloG2020, karthikaReviewClassicalNonclassical2016b} . CNT proceeds by projecting the full infinite-dimensional field dynamics onto a single slow reaction coordinate, the radius $R$ of a spherical nucleating droplet. 
Assuming that the droplet profile is quasi-static and spherically symmetric, the field dynamics typically reduces to a one-dimensional stochastic process of the form~\footnote{Note that in the LDT limit of $T\to 0$, the It\^o and Stratanovich interpretations of \eqref{eq:Rdot_intro} coincide, despite the presence of multiplicative noise.}
\begin{align}\label{eq:Rdot_intro}
    \dot{R} = \mathcal{M}(R)\left[-\partial_R U(R)\right] + \sqrt{2T\mathcal{M}(R)}\,\eta(t),
\end{align}
where $U(R)$ is the effective free energy (or quasi-potential) as a function of radius, $\mathcal{M}(R)$ is an effective mobility, and $\eta(t)$ is a Gaussian white noise with $\langle \eta(t)\eta(t') \rangle = \delta(t-t')$. 
The nucleation rate is then determined by the height of the barrier $U(R_\mathrm{c})$ at the critical radius $R_\mathrm{c}$, with $\mathcal{P} \asymp \exp(-U(R_\mathrm{c})/T)$ equivalent to a Kramers' escape rate problem~\cite{RevModPhys.62.251}.
This procedure applies in equilibrium to both conserved (Model B) and non-conserved (Model A) dynamics~\cite{brayTheoryPhaseorderingKinetics, Lutsko2012Jan}. 

Extending CNT to non-equilibrium systems is considerably more challenging. The task is important, because CNT-like precepts have increasingly (if informally) been used in active systems, both at fixed particle number~\cite{richard2016nucleation,redner2016classical,levis2017active} and in subcellular liquid-liquid phase separation involving chemically reacting species~\cite{cho2023tuning,shimobayashiNucleationLandscapeBiomolecular2021b}, and also in the context of population dynamics \cite{michaelsNucleationFrameworkTransition2020, Gandhi1999Sep, tanakaSpatialGeneDrives2017}. Apart from some special cases where there is effectively a mapping onto an equilibrium problem~\cite{Ziethen2024Jun,noahprl}, these applications lack fundamental justification.

Recently however, CNT was successfully extended to address
active phase separating systems with a conserved order parameter~\cite{catesClassicalNucleationTheory2023}. In that case, the interfacial density profile of a nucleating droplet is close enough to that of a droplet relaxing back towards equilibrium that the difference can be ignored at CNT level. (This point is explained in detail later in this paper; see Sec.~\ref{sec:AMB+}.) Crucially, as we shall see, the same is not true for the non-conserved case, making it significantly more subtle. 

In a short companion article~\cite{prl}, we address this gap by reporting a consistent projection of the field dynamics onto the radial coordinate for non-equilibrium non-conserved field theories. In the resulting Nonequilibrium Nucleation Theory (NNT), active terms that are purely deterministic at field level, once projected into the Langevin equation~\eqref{eq:Rdot_intro} for the reaction coordinate,
modify not just its deterministic
dynamics {\em but also 
its noise structure}. This means that breakdown of the TRR ansatz has deep consequences for the resulting quasipotential landscape.

Since the word `Classical' in the term `Classical Nucleation Theory' could refer (at least in part) to its equilibrium origins, we propose that this nomenclature be restricted to theories that do assume the TRR ansatz \footnote{Notice, however, that the term non-classical nucleation theory was widely employed in equilibrium systems to denote cases where the system visits states corresponding to non-spherical liquid droplets along the nucleation dynamics.}. 
NNT, which makes no such ansatz, then reduces to CNT only when TRR is actually valid to the order required by the theory, as holds for a conserved order parameter (see Sec.~\ref{sec:AMB+}).

The current paper provides the full technical account of the NNT summarized in~\cite{prl}. It also unifies the approach taken there (based on explicit derivation of~\eqref{eq:Rdot_intro}) with a direct approach based on action minimization. In Sec.~\ref{sec:general-theory} we introduce the general class of models that we consider in this paper. We begin by describing how to define the radius of a nucleating droplet and describing the stochastic motion of interfaces in one-dimensional systems. After recapping the derivation of NNT via the stochastic route as described in~\cite{prl}, we show how to retrieve NNT by working at the level of the action (action route), and show that lifting the TRR ansatz substantially changes the nucleation barrier. In Sec.~\ref{sec:ama} and \ref{sec:pd}, we apply the framework to two concrete non-equilibrium models: An out-of-equilibrium generalization of Model A (in Halperin-Hohenberg classification)~\cite{Allen1979Jun} and a population dynamics model. (The latter has two choices of noise, each breaking the fluctuation-dissipation relation that would hold in equilibrium.)
In both cases, we derive closed-form analytical expressions for the quasi-potential and nucleation barrier, making explicit how non-equilibrium activity shifts the quasi-potentials. Furthermore, we provide a straightforward derivation the dynamics of capillary waves and confirm their stability such that the nucleation dynamics is unaffected by them. 
In Sec.~\ref{sec:num}, we validate our analytical predictions for AMA and population dynamics against numerical action minimization, and illustrate the geometry of the instanton trajectories in each system.
Finally, in Sec.~\ref{sec:AMB+} we show how our general NNT theory can encompass known results for the phase separation of a conserved order parameter~\cite{catesClassicalNucleationTheory2023}, and confirm the applicability of the TRR anzatz in this case. We conclude in Sec.~\ref{sec:discussion} with a short discussion.

\section{General theory}\label{sec:general-theory}
CNT describes a nucleating spherical droplet that, due to (thermal) fluctuations, 
reaches a
critical radius $R_\mathrm{c}$ and then invades the full system deterministically. The existence of a critical radius is associated with the competition of an invading velocity $v_0$ of the flat interface (which, in equilibrium, is associated with a mismatch of the free energy in the two bulk phases) with a surface tension that causes the curved droplet to shrink~\cite{brayTheoryPhaseorderingKinetics}. As we will see, in non-equilibrium contexts, both $v_0$ and the surface tension are modified by activity. Nonetheless, we follow a standard precept of CNT (as discussed in~\cite{prl}) in assuming that $v_0$ is a small parameter, which (at fixed tension) also means that $R_\mathrm{c}$ is large. 

In this Section we develop a general theory of nucleation in nonconserved non-equilibrium field theories. In Sec.~\ref{sec:models-cn} we introduce the class of models considered. In Sec.~\ref{sec:int-position} we discuss how to identify the radius of the nucleating droplet: This is a crucial technical step in order to derive NNT without the need to {\em explicitly} compute the deviation of the instanton trajectory from the time-reversed relaxation (TRR) dynamics~\cite{prl}. En route, we discuss in Sec.~\ref{sec:interface-1d} the stochastic dynamics of one-dimensional interfaces.
In Sec.~\ref{sec:interface-drop}, we then recap the derivation of NNT via the stochastic route presented in~\cite{prl}. We show in Sec.~\ref{sec:action-trs} that neglecting the difference between the instanton trajectory and the TRR one leads to incorrect results. We provide an alternative derivation of NNT working at the level of the Freidlin-Wentzell action in Sec.~\ref{sec:action-non-trs}. 

Throughout this Section we assume interfacial stability so that the nucleating droplet remains spherical.

\subsection{Models}\label{sec:models-cn}
\begin{figure}[t]
    \centering
\includegraphics[width=\columnwidth]{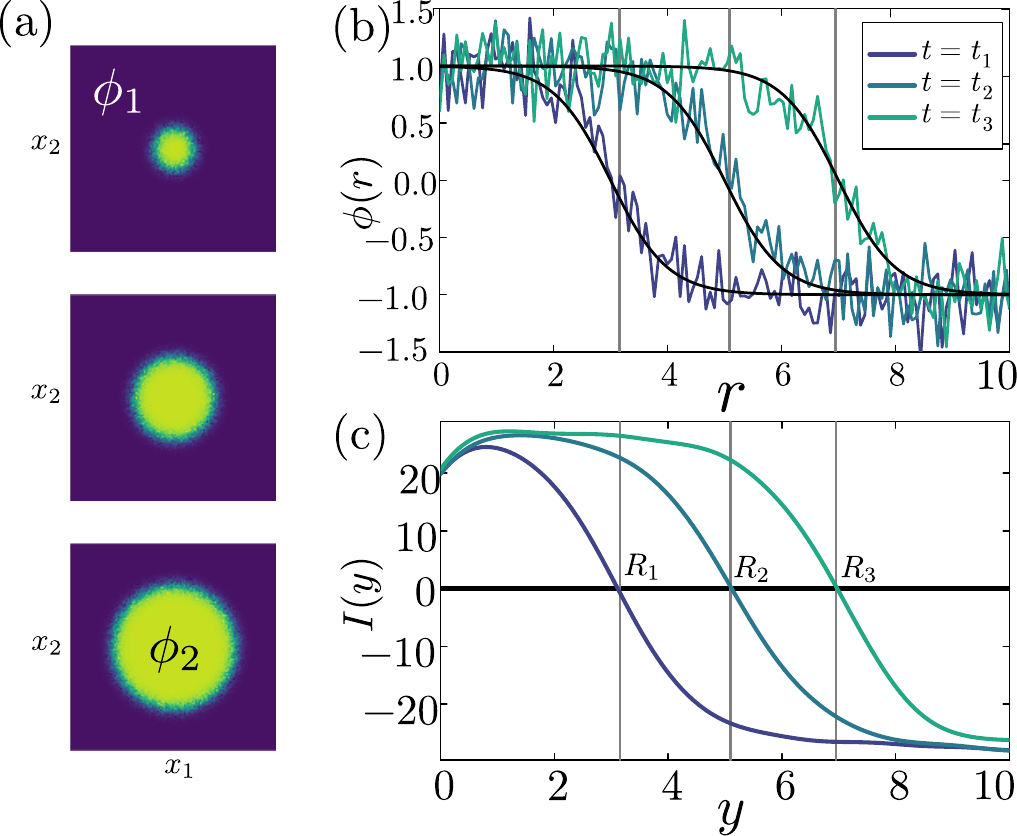}
    \caption{(a) Nucleation of a two-dimensional spherical droplet; after reaching the critical radius, the $\phi_2$ phase invades the remainder by a deterministic relaxation. (b) Shape of the density profile as a function of the radial coordinate (coloured lines) while the black solid lines correspond to the deterministic ($T=0$) $\varphi(r)$ profiles. (c) Plot of $I(y)$ defined in Eq.~\eqref{eq:defn}; The root $I(R)=0$ defines the radius $R$ of the droplet.}
    \label{fig:fig1}
\end{figure}

We consider models where the global order parameter $\phi_0=\int d\p{x} \,\phi$ is not dynamically conserved. We use the notation
\begin{align}\label{eq:genmodel}
    \partial_t \phi = - \mu[\phi] + \sqrt{2T D(\phi)} \xi
\end{align}
where $\xi$ is unit Gaussian white noise such that $\langle \xi(\p{x},t) \rangle = 0$ and $\langle \xi(\p{x},t) \xi(\p{x}',t') \rangle = \delta(\p{x}-\p{x}') \delta(t-t')$. 
We assume that the deterministic drift $\mu[\phi]$ in \Eqref{eq:genmodel} admits two locally stable fixed points $\phi_{1,2}$ of which we assume $\phi_1$ to be metastable and $\phi_2$ to be globally stable. Thus a droplet of phase $\phi_2$ grows in a surrounding $\phi_1$ phase until it fully invades it (see Fig.~\ref{fig:fig1}(a)). Although our results could be extended to general functionals $\mu[\phi]$, we will assume here that $\mu$ is a local function of $\phi$ and its derivatives, while $D(\phi)>0$ is a local function of $\phi$ only. To keep our notation compact, we express $\mu$ (loosely called the ``chemical potential" below) as $\mu[\phi]=\mu_0[\phi]+r^{-1}\mu_1[\phi] + \cdots$. Here $\mu_{0}, \mu_{1}$ depend on $r$ only via $\phi$, and the ellipsis denotes contributions proportional to $r^{-n}$ with $n\geq 2$.

Examples of Eq.~\eqref{eq:genmodel} include equilibrium Model A ~\cite{hohenbergTheoryDynamicCritical1977,brayTheoryPhaseorderingKinetics,onuki2002phase}), which corresponds to $\mu=\delta\mathcal{F}/\delta\phi$ with $\mathcal{F}[\phi]= \int (f(\phi)+K |\nabla\phi|^2/2)\,d{\bf x}$, $D=1$, and $f(\phi)$ a double-well local free energy. 
Importantly, we also include non-equilibrium models where $\mu[\phi]/D(\phi)$ is not the functional derivative of any $\mathcal{F}[\phi]$; see Sec. \ref{sec:ama} and \ref{sec:pd} for examples.

\subsection{Definition of the interface position}\label{sec:int-position}
A crucial technical step is to choose carefully the definition of the interfacial position and hence the droplet radius (reaction coordinate). We follow ideas that were introduced in the past in the context of reaction-diffusion systems~\cite{kuramoto1980instability,kawasaki1982kinetic} and to describe capillary waves in equilibrium models A~\cite{kawasaki1982kinetic-i} and C~\cite{bausch1991effects}.

We first write
\begin{align}\label{eq:fielddev}
\phi(\p{x},t) = \varphi(r,R) + \epsilon(\p{x},t)
\end{align}
where $\varphi$ is the interfacial field profile during the relaxation dynamics of a spherical droplet at radius $R$ (obeying ~\eqref{eq:genmodel} with $T=0$), and $\epsilon$ takes into account all deviations from this. (This $\epsilon$ will be small, but not negligible, in the NNT limit of small $v_0$.) We next introduce the notation 
\begin{align}\label{eq:varphi-decomposed}
\varphi(r,R) = \varphi_0(r-R)+\frac{1}{R}\varphi_1(r-R)+\mathcal{O}(1/R^2)\,,
\end{align}
in which $\varphi_0$ is the density profile of a flat interface, and $\varphi_1$ is the first correction due to curvature.
Finally we introduce a function $\psi(y)$ whose $y$-derivative $\psi'$ is peaked at $y=0$, and whose precise form will be chosen later. We consider
\begin{align}\label{eq:defn}
    I(y) := \int_{\p{x}} \, \epsilon(\p{x},r,t) \psi'(r-y) ,
\end{align}
where $\int_\p{x}=\int d\p{x}$ is the volume integral, and $r=|\p{x}|$ is the radial coordinate. We define the radius of the droplet $R$ as the value such that $I(R)=0$ (see Fig.~\ref{fig:fig1}(c)). 

The existence of the required root is demonstrable when $\epsilon$ is small. Taking the $y$-derivative of Eq.~\eqref{eq:defn}, we have 
\begin{align}\label{eq:py-Iy}
    \partial_y I(y) = \int_\p{x} \, \psi' \varphi' + \mathcal{O}(\epsilon) \,
\end{align}
In all examples considered in this paper,~\eqref{eq:py-Iy} implies that $I(y)$ is monotonic, and hence a root $I(R)=0$ exists.

We can use a similar definition for the position $X(t)$ for an interface moving in one dimension. This obeys
\begin{align}\label{eq:def-1d-interface}
    0=\int_x \, \epsilon_{1d}(x,X(t),t) \psi'(x-X(t))
\end{align}
where now $\epsilon_{1d}(x,X(t),t)=\phi(x,t)-\varphi_0(x-X(t))$ and $\varphi_0$ solves Eq.~\eqref{eq:genmodel} with $T=0$ in $d=1$. 
We will next use this to derive the interface velocity and the fluctuations of its position in the small noise regime. (The idea will be used again to consider the dynamics of capillary waves in Sec.~\ref{subsec:cap_ama} and \ref{sec:pop-dyn-cw} below.)

\subsection{Motion of a one-dimensional (flat) interface}\label{sec:interface-1d}
We now exemplify the use of the above definitions by deriving the motion of a one-dimensional interface under conditions of
small deterministic drift $v_0$ and weak noise $T$.
Taking the time derivative of Eq.~\eqref{eq:def-1d-interface} we find
\begin{align}
    \dot X &= -  \frac{\int_x\dot \phi \, \psi'}{\int_x \psi' \varphi_0' - \int_x \psi'' \epsilon_{1d}}\nonumber\\
    &=\frac{\int dx \mu_0[\phi]\psi'}{\int_x \psi' \varphi_0' - \int_x \psi'' \epsilon_{1d}} - \frac{\sqrt{2T}\int_{x} \, \psi' \sqrt{ D(\phi)} \xi}{\int_{x} \, \psi' \varphi_0'- \int_{x} \psi'' \epsilon_{1d}}\;.
\end{align}
We next expand $\mu_0[\phi]= \mu_0[\varphi_0 + \epsilon_{1d}]=\mu_0[\varphi_0] + \Lmu \epsilon_{1d} + \mathcal{O}(\epsilon_{1d}^2)$, and $D(\phi)=D(\varphi_0) + \partial_\phi D (\varphi_0)\epsilon_{1d} + \mathcal{O}(\epsilon_{1d}^2)$. $\mathcal{L}$ is a linear operator found by expanding around $\varphi_0$. In the regime of small $\epsilon_{1d}$ we obtain
\begin{align}\label{eq:1d-interface-int}
    \dot X =v_0 +\frac{\int_x \psi' \Lmu \epsilon_{1d}}{\int_x \psi'\varphi_0'}- \frac{\sqrt{2T}\int_{x} \, \psi' \sqrt{ D(\varphi_0)} \xi}{\int_{x} \, \psi' \varphi_0'}+\mathcal{O}(...)
\end{align}
where
\begin{align}\label{eq:v0}
    v_0 = \frac{\int_x  \mu_0[\varphi_0] \psi'}{\int_x \psi' \varphi_0'}\,,
\end{align}
and $\mathcal{O}(...)=\mathcal{O}(\epsilon_{1d}^2,\epsilon_{1d} \sqrt{T},v_0\epsilon_{1d})$.

We may now make a choice of $\psi'$ to simplify \eqref{eq:1d-interface-int} dramatically, leading to a closed equation for $X(t)$. Crucially, we expect $\epsilon_{1d}\sim \mathcal{O}(\sqrt{T})$ so that, for a generic choice of $\psi'$, the second term in Eq.~\eqref{eq:1d-interface-int} must be explicitly computed in order to obtain the correct noise amplitude in the equation for $X(t)$. However, this can be completely avoided by choosing $\psi'\in\mathrm{Ker(\mathcal{L}^\dagger)}$~\cite{kuramoto1980instability,kawasaki1982kinetic,kawasaki1982kinetic-i,bausch1991effects}.
Under this choice, $\int_x \psi'\mathcal{L}\epsilon_{1d}=0$ and hence we obtain \begin{align}\label{eq:1d-xdot}
    \dot X = v_0 + \sqrt{2T\mathcal{M}_{1d}}\,\eta(t) +\mathcal{O}(T,v_0\sqrt{T})\;,
\end{align}
with white noise $\langle\eta(t)\eta(t')\rangle=\delta(t-t')$.
Here the interfacial drift velocity $v_0$ obeys~\eqref{eq:v0}, and the mobility is  
\begin{align}\label{eq:1d-mob}
    \mathcal{M}_{1d} = \frac{\int_x \, \psi'^2 D(\varphi_0) }{ \left[\int_x \, \psi' \varphi_0'\right]^2}\;.
\end{align}
The key idea is that by making the choice $\psi'\in\mathrm{Ker(\mathcal{L}^\dagger)}$, corrections due to the deviations of the density profile from the deterministic one (encoded in $\epsilon_{1d}$) contribute only at subleading order in Eq.~\eqref{eq:1d-xdot}, whereas for generic $\psi'$, this would not be true.

We emphasize that the choice of $\psi'$ does not change the deterministic drift $v_0$. While Eq.~\eqref{eq:v0} seems to depend on $\psi'$, any other choice leads to the same drift velocity $v_0$ Indeed, in the absence of noise ($T=0$), $\epsilon_{1d}=0$ at all times. This follows from the definition of $\varphi_0$, which is the solution to Eq.~\eqref{eq:genmodel} with $T=0$ in $d=1$, and implies that the choice of $\psi'$ is irrelevant. This fact will be fully discussed in Sec.~\ref{sec:AMA-1d-interface} for a specific example.

\subsection{NNT: Stochastic route}\label{sec:interface-drop}
As shown in the companion article~\cite{prl}, the Langevin dynamics of a reaction coordinate $R$, comprising the radius of a spherical droplet in $d\geq 2$ dimensions can be derived directly. We call this approach `the stochastic route' and here 
briefly restate the main results. Taking the time derivative of Eq.~\eqref{eq:defn}, and choosing $\psi'\in \mathrm{Ker}(\mathcal{L}^\dagger)$
we find that, to leading order in large $R$ and $R_\mathrm{c}$ (or, equivalently, small $v_0$), the projected dynamics becomes
\begin{align}
    \dot R=v_0 + \frac{\int_r( \, \psi'\mu_1[\varphi_0]+\psi'\mathcal{L}\varphi_1)}{R\int_r \, \psi'\varphi_0' }- \frac{\sqrt{2T}\int_\p{x} \, \psi'\sqrt{ D(\varphi_0)} \xi}{\int_\p{x}  \, \psi' \varphi_0'}
    \label{eq:gen_Rdot_drop}
\end{align}
where $v_0$ obeys \eqref{eq:v0}. Note that the term containing $\mathcal{L}$ exactly vanishes when $\psi'\in\mathrm{Ker}(\mathcal{L}^\dagger)$ but we kept it here to aid later comparisons. 
By recasting (\ref{eq:gen_Rdot_drop}) into the form of Eq.~(\ref{eq:Rdot_intro}), We can identify the interface mobility 
\begin{align}
\mathcal{M}(R)= 
     \frac{\int_r \, \psi'^2 D(\varphi_0) }{R^{d-1} S_d\left[\int_r \, \psi' \varphi_0'\right]^2}\,,
     \label{eq:mob-gen}
\end{align}
and the quasipotential
\begin{align}\label{eq:URgen}
    U(R) =    \frac{S_d R^{d-1}\sigma  
    \int_r \, \psi' \varphi_0' }{\int_r \, \psi'^2 D(\varphi_0)}\left( 1 - \frac{R (d-1)  }{R_\mathrm{c} d } \right)\,,
\end{align}
where $S_d$ is the $d$-dimensional solid angle.
Here, the critical radius is given by
\begin{align}\label{eq:rc-gen}
    R_\mathrm{c} = \frac{(d-1)\sigma}{ \int_r \, \psi' \mu_0[\varphi_0]},
\end{align}
with $\sigma = -\int_r \, \psi'  \mu_1[\varphi_0]/(d-1)$ is the interfacial tension~\cite{prl}.
We show in Sec.~\ref{sec:action-non-trs} how the same results can also be recovered by minimizing the action.

Similarly to the flat interface equation~\eqref{eq:1d-interface-int}, the deterministic drift of the radius is correctly found by Eq.~\eqref{eq:gen_Rdot_drop} for any $\psi'$ even if $\psi'\notin \textrm{Ker}(\mathcal{L}^\dagger)$. However, as shown in~\cite{prl} and confirmed by other means below, the mobility (and hence the noise amplitude) in Eq.~\eqref{eq:gen_Rdot_drop} is incorrectly determined unless the choice $\psi'\in \textrm{Ker}(\mathcal{L}^\dagger)$ is made. 

\subsection{Time Reversed Relaxation: Action route} \label{sec:action-trs}
In conserved phase-separating active systems, the effective dynamics for $\dot{R}$ describing nucleation can be obtained by assuming that the instanton is the time-reversal of the relaxational path, see~\cite{catesClassicalNucleationTheory2023} and Sec.~\ref{sec:AMB+}. This requires that for any instantaneous droplet radius $R$ during the nucleation process, the radial density profile $\phi(r,R)$ is the same as the one, $\varphi(r,R)$, arising when passing through that radius via relaxation (meaning, the noiseless dynamics back to the metastable state at $\phi_1$).
We demonstrate in this Section that this does not hold for nonconserved non-equilibrium systems. 

The FW action associated with the stochastic dynamics of the full field $\phi(\p{x},t)$ in Eq.~\eqref{eq:genmodel} is given by~\cite{Freidlin1998}
\begin{align}\label{eq:FWgen}
    \mathcal{A} = \frac{1}{4T}\int_{\p{x},t}  \frac{(\partial_t \phi +  \mu[\phi])^2}{ D(\phi)}\;.
\end{align}
Imposing the TRR ansatz that $\phi=\varphi(r,R)$ is the relaxation profile, we obtain 
\begin{align}
    \mathcal{A} = \frac{1}{4T}\int_{\p{x},t} \left(\dot R \partial_r \varphi - \mu[\varphi]+\mathcal{O}(\dot{R}/R^2)\right)^2 D^{-1}(\varphi)\;.
\end{align}
After some further algebra, this becomes
\begin{align}\label{eq:action-TRS}
&\mathcal{A}=\nonumber\\
    &\frac{1}{4T}\int_t\Bigg[ \left(\int_\p{x}\frac{\varphi'^2}{D}\right) \left(\dot R- \left( \int_r\frac{\varphi'^2}{D}\right)^{-1}\left(\int_r \frac{\mu \varphi'}{D}\right) \right)^2
    \Bigg]
    \nonumber\\
    & - \frac{1}{4T}\int_t \left[\left(\int_\p{x}  \frac{\mu\varphi'}{D}\right)^2\left(\int_\p{x} \frac{\varphi'^2}{D}\right)^{-1} -  \int_\p{x}   \frac{\mu^2}{D}\right]\nonumber\\
    &+\mathcal{O}(\dot{R}/R^3,v_0 \dot{R}/R^2)\,.
\end{align}
Given that $\varphi$ is the relaxational density profile, we have $\mu[\varphi]=v_\mathrm{R} \varphi'$, where $v_\mathrm{R} =  \left(\int_r \frac{\mu[\varphi] \varphi'}{D(\varphi)}\right)/\left( \int_r\frac{\varphi'^2}{D(\varphi)}\right)$ is the relaxational velocity of a droplet of radius $R$. Using this identity, the second term $\int_t\left[...\right]$ in Eq.~\eqref{eq:action-TRS} exactly vanishes. The remaining action leads directly to
the following Langevin equation, suppressing subleading terms:
\begin{align}
    \dot R &= \left( \int_r\frac{\varphi'^2}{D}\right)^{-1}\left(\int_r \frac{\mu \varphi'}{D}\right) + \sqrt{2T\left(\int_\p{x}\frac{\varphi'^2}{D} \right)^{-1}}\eta(t),
\end{align}
Using \eqref{eq:varphi-decomposed} and expanding for small $v_0$ and large radius, we have,
\begin{align}
    \dot R &=  \left( \int_\p{x}\frac{\varphi_0'^2}{D}\right)^{-1} \int_\p{x} \left[\frac{\mu_0\varphi'_0}{D}+\frac{\varphi_0'\mathcal{L}\varphi_1+\varphi_0'\mu_1[\varphi_0]}{R \, D}\right]\nonumber\\
    &+ \sqrt{2T\left(\int_\p{x}\frac{\varphi_0'^2}{D} \right)^{-1}}\eta(t) \;, \label{eq:rdot-TRR}
\end{align}
with terms neglected $\mathcal{O}(R^{-2},v_0 R^{-1}, \sqrt{T}R^{-(d+1)/2})$.

In conclusion, by assuming the TRR ansatz, we obtain Eq.~\eqref{eq:Rdot_intro} with $U=U_\varphi$, $\mathcal{M}=\mathcal{M}_\varphi$, where 
\begin{align}\label{eq:Uphi}
U_\varphi &:=\frac{-S_d R^{d-1} \int_r \frac{\varphi_0'(\mathcal{L}\varphi_1+ \mu_1 )}{D}}{(d-1)}\left( 1 - \frac{R (d-1)  }{R_\mathrm{c} d } \right),
\\
\mathcal{M}_\varphi^{-1} &:= R^{d-1}S_d\int_r (\varphi_0'^2/D)\;,
\end{align}
from which the critical radius follows as
\begin{align}\label{eq:Rc-TRS}
    R_\mathrm{c,\varphi} = -\frac{\int_\p{x}\frac{\varphi_0'\mathcal{L}\varphi_1+\varphi_0'\mu_1[\varphi_0]}{ D}}{\int_\p{x} \frac{\mu_0\varphi'_0}{D}}\;.
\end{align}

Of these three results, only the critical radius $R_\mathrm{c}$ is correct. (This is because it locates the unstable fixed point of the deterministic dymamics which can be found without considering the noise.) The mobility, and crucially the quasipotential and with it the barrier height, are not given correctly by the TRR ansatz and indeed they contradict the results (\ref{eq:mob-gen},\ref{eq:URgen}) of the stochastic calculation which carefully avoided that ansatz. The same incorrect results can however be replicated by a less careful version of the stochastic route in which
one chooses $\psi'=\varphi_0'/D$ (which is not $\in\mathrm{Ker(\mathcal{L}^\dagger)}$) and yet discards the $\epsilon$ term in \eqref{eq:gen_Rdot_drop} (which therefore does not vanish). 

\subsection{NNT: Action route}\label{sec:action-non-trs}
In this Section we show that a self-consistent minimization of the FW action, to the required order in a $1/R$ expansion and avoiding the TRR ansatz, recovers the same Nonequilibrium Nucleation Theory as was derived in~\cite{prl} and reviewed above, namely Eqs. (\ref{eq:Rdot_intro},\ref{eq:mob-gen},\ref{eq:URgen},\ref{eq:rc-gen}). 
We perform here the calculation for the special case $D=1$ although the general case can be obtained with the same argument. 

We start from the FW action in Eq. \eqref{eq:FWgen} and decompose $\phi$ as
\begin{align}
    \phi(\mathbf{x}, t) =\varphi(r,R)+\epsilon_\mathrm{A}(\p{x},R,t) \;.
\end{align}
We introduce the subscript A on $\epsilon$ to emphasize that this is not mathematically the same object as the deviation in \eqref{eq:fielddev}. (Indeed, in LDT the FW action is evaluated on smooth paths whereas the Langevin trajectories in \eqref{eq:genmodel} are not smooth.)

Assuming the action to be dominated by paths for which $\epsilon_\mathrm{A}$ is small, we obtain $\mathcal{A} = \frac{1}{4T}\int_{\p{x},t} \vartheta^2$, 
where 
\begin{align}\label{eq:gen-vartheta-def-cn}
\vartheta=-\dot R\varphi' + \mu[\varphi] + (\partial_t +\mathcal{L}_\varphi)\epsilon_\mathrm{A}+\mathcal{O}(\epsilon_\mathrm{A}^2)\,.
\end{align}
Minimizing this with respect to $\epsilon_\mathrm{A}$ leads to 
\begin{align}\label{eq:EL-moment}
    (-\partial_t + \mathcal{L}_\varphi^\dagger)\vartheta = 0\,.
\end{align}
To solve \eqref{eq:EL-moment}, we first neglect $\partial_t \vartheta$ and then show {\em a posteriori} that this neglect is justified. 

The solution to $\mathcal{L}_\varphi^\dagger \vartheta=0$ has the form $\vartheta = c \psi'$ where $\psi'\in \mathrm{Ker}(\mathcal{L}_\varphi^\dagger)$.
The unknown constant $c$ can be determined by multiplying by $\psi'$ and integrating over the entire space
\begin{align}
    \int_\p{x} \vartheta \psi' d\mathrm{r} = c \int_\p{x} \psi'^2 
\end{align}
and consequently
\begin{align}
    c=\frac{-\dot R \int_\p{x} \varphi' \psi'  + \int_\p{x} \mu[\varphi] \psi' } {\int_\p{x}\psi'^2 }+\mathcal{O}(\partial_t \epsilon_\mathrm{A}, \epsilon_\mathrm{A}^2,\partial_t\vartheta)
\end{align}
where we assumed that $\partial_t \epsilon_\mathrm{A}$ is small. Further using that $\vartheta$ is proportional to $\psi'$ and hence sharply peaked at the interface, we obtain
\begin{align}\label{eq:gen-vartheta-cn}
    \vartheta &= \frac{-\dot R \int_r \varphi_0' \psi' + \int_r \mu_0[\varphi_0] \psi' + R^{-1} \int_r \mu_1[\varphi_0]\psi'} {\int_r\psi'^2 }\psi'\nonumber\\
    &+\mathcal{O}( 1/R^2,\epsilon_\mathrm{A}^2,\partial_t\epsilon_\mathrm{A},\partial_t\vartheta)\,.
\end{align}
It should be noticed that $\psi'$ was defined here as $\psi'\in \mathrm{Ker}(\mathcal{L}_\varphi^\dagger)$ where $\mathcal{L}_\varphi^\dagger$ is the linear operator corresponding to $\mu$. In the stochastic route, $\psi'$ was instead defined with respect to the linear operator $\mathcal{L}^\dagger$ associated with $\mu_0$. 
Because $\vartheta$ is proportional to $\psi'$, which is sharply peaked at the interface, only $\vartheta$ evaluated at $r=R$ matters for the action to the order we are interested in. As such, 
the difference between kernels is in terms of order $\epsilon_\mathrm{A}/R$ which are negligible. Injecting eq. \eqref{eq:gen-vartheta-cn} in the action, we obtain
\begin{align}
    \label{eq:actionleading} \mathcal{A} &= \frac{1}{4T}\int_t \frac{R^{d-1} S_d \left(\int_r \varphi'_0 \psi' \right)^2}{\int_r\psi'^2 }\\ 
    &\times\left(\dot R - \frac{\int_r \mu_0[\varphi_0] \psi'+R^{-1}\int_r\mu_1[\varphi_0]\psi'}{\int_r \varphi'_0 \psi'} +\mathcal{O}(...)\right)^2,\nonumber
\end{align}
where $\mathcal{O}(...)=\mathcal{O}(\epsilon_\mathrm{A}^2,\partial_t\epsilon_\mathrm{A},\epsilon_\mathrm{A}/R,1/R^2,\partial_t\vartheta,v_0/R)$. To the required order, this is precisely the action associated with the $\dot{R}$ equation for the reaction coordinate, in (\ref{eq:Rdot_intro},\ref{eq:mob-gen},\ref{eq:URgen},\ref{eq:rc-gen}) obtained via the stochastic route. 

We are left to check the self-consistency of the assumption that the $\mathcal{O}(...)$ terms \eqref{sec:interface-drop} can indeed be neglected. We need concern ourselves with these terms only for $r\simeq R$. Still neglecting $\partial_t\epsilon_\mathrm{A}$ for the moment, we conclude from Eq.~\eqref{eq:gen-vartheta-def-cn} and ~\eqref{eq:gen-vartheta-cn} that $\epsilon_\mathrm{A} \sim \mathcal{O}(\dot R, v_0,1/R)$. Leveraging on this result we also conclude from Eqs.~\eqref{eq:gen-vartheta-def-cn} and ~\eqref{eq:gen-vartheta-cn} that
$\partial_t \epsilon_\mathrm{A} \sim \mathcal{O}(\ddot R, \dot R^2, \dot R v_0,\dot{R}/R,1/R^2)$.
From these results and Eq.~\eqref{eq:gen-vartheta-cn} we further have that
$\partial_t \vartheta\sim \mathcal{O}(\ddot R, \dot R^2, \dot R v_0,\dot{R}/R,1/R^2)$. The self-consistency argument is concluded by noticing that, for droplets on the instanton pathway, $\dot{R}\sim \mathcal{O}(1/R,v_0)$ and $\ddot{R}\sim\mathcal{O}(\dot{R}/R^2)$. This finally confirms that all terms in $\mathcal{O}(...)$ are negligible. Notice also that this argument allows to conclude that $\epsilon_\mathrm{A}\sim\mathcal{O}(v_0)$ insofar as both $R$ and $R_\mathrm{c}$ are large, a fact that we will check numerically in Sec.~\ref{sec:num} below.

\begin{figure*}[t]
    \centering
\includegraphics[width=0.9\textwidth]{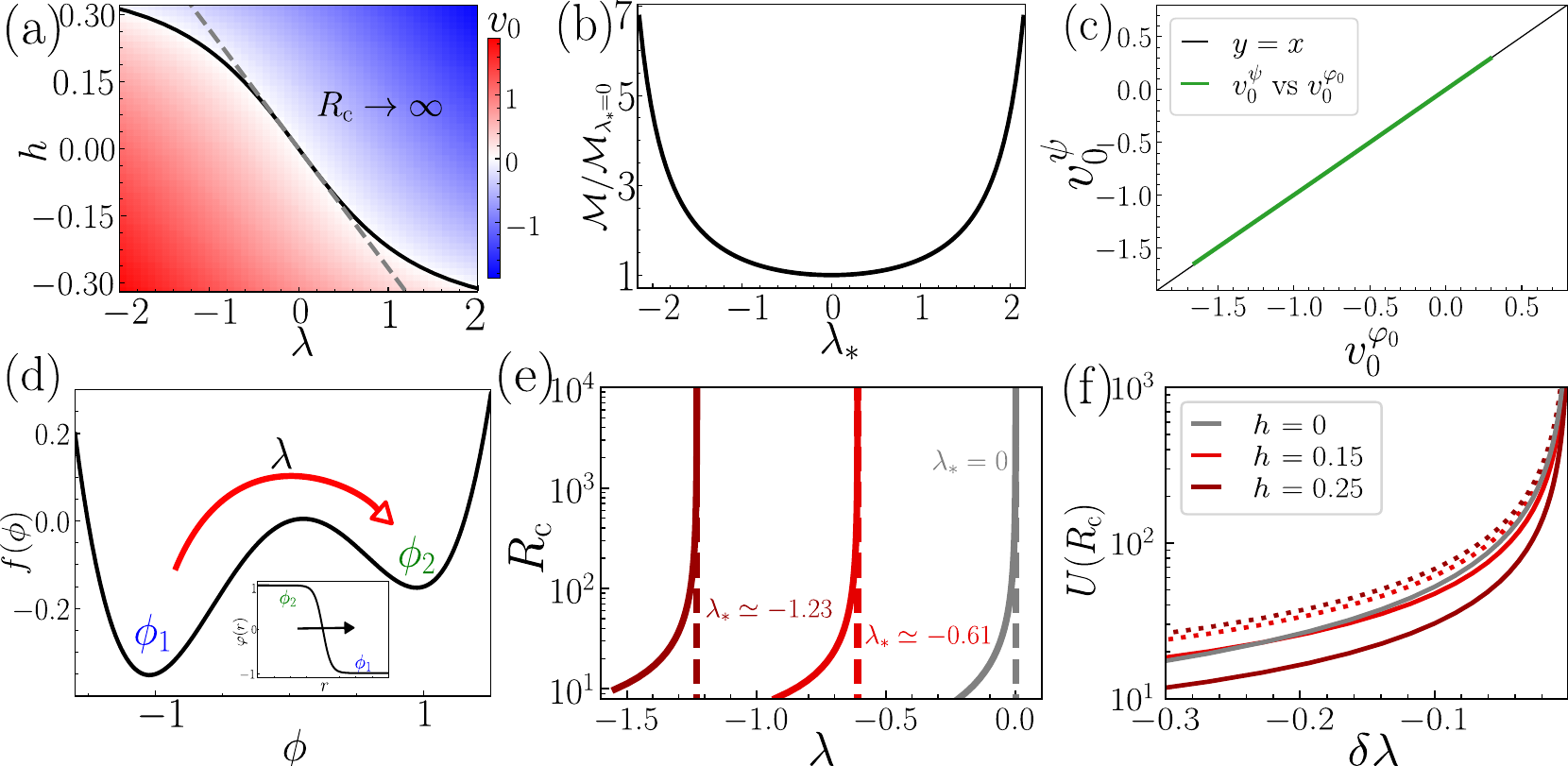}
    \caption{(a) The velocity of one-dimensional interface ($v_0$) as a function of $h,\lambda$ obtained from numerical integration of AMA via a pseudo-spectral method. For the simulations the number of points for space discretization is $N_x=1024$, domain length $L=16\pi$ and $dt = 10^{-3}$. The grey line corresponds to the perturbative result $h=-4\lambda/15$ for $h,\lambda \ll1$ [see Eq.~\eqref{eq:pertrc}]. The solid black line is where the interface is static. (b) The normalized mobility of the interface along the static line in (a). (c) Numerical comparison of $v_0^\psi$ and $v_0^{\varphi_0}$ for $h=0.25$, found via a pseudo-spectral method.  (d) Example plot of the potential $f(\phi)$ for $h=0.1$. At large enough $\lambda$, stability is reversed so that the globally stable (nucleating) phase is the one of higher $f$.
    (e) The critical radius $R_\mathrm{c}(\lambda)$ for three values of $h=(0,\, 0.15,\, 0.25)$, where the dashed lines indicate the divergence of $R_\mathrm{c}$ at $\lambda = \lambda_*(h)$. (f) The barrier height for $\lambda=\lambda_*(h)+\delta \lambda$, where the solid line corresponds to the NNT quasipotential $U(R_\mathrm{c})$ and the dotted line is $U_\varphi(R_\mathrm{c})$ found by the TRR ansatz. All the results correspond to the case of $K=1$.}
    \label{fig:ama_together}
\end{figure*}

\section{Active Model A}\label{sec:ama}
We now apply our NNT formalism to Active Model A (AMA), a minimal field theory describing phase-ordering in active systems~\cite{PhysRevLett.124.240604}. After introducing the model, we show in Sec.~\ref{sec:AMA-ker} that the kernel of $\mathcal{L}^\dagger$ can be explicitly found. In Sec.~\ref{sec:AMA-1d-interface} we discuss the stochastic dynamics of one-dimensional interfaces. 
In Sec.~\ref{sec:NNT-AMA}, we explicitly obtain predictions for nucleation from NNT and show that CNT incorrectly predicts nucleation rates. In the regime close to equilibrium, we further discuss explcit perturbative results in Sec.~\ref{sec:AMA-near-eq}. In Sec.~\ref{subsec:cap_ama}, we discuss capillary fluctuations of the interface, showing that deformations of the droplet shape are of order $\mathcal{O}(T/\sigma_\mathrm{cw})$, where $\sigma_\mathrm{cw}>0$ is the capillary-wave interfacial tension, and are thus negligible as expected in the nucleation dynamics. 

\subsection{Model and $\mathrm{Ker}(L^\dagger)$}\label{sec:AMA-ker}
AMA is defined by Eq.~\eqref{eq:genmodel} with $D=1$,
\begin{align}\label{eq:ama}
\mu[\phi] = \partial_\phi f(\phi) - K \nabla^2 \phi + \lambda (\nabla\phi)^2
\end{align}
and $f(\phi) = -\phi^2/2+\phi^4/4+h\phi$. Here, $h$ plays the role of an external field. 
The stable homogeneous states $\phi_{1,2}$ are solutions to $\partial_{\phi}f(\phi)=0$. 
As shown below, which one among $\phi_{1,2}$ is the stable and metastable state depends both on $h$ and $\lambda$. 
Note that AMA breaks detailed balance because Eq. \eqref{eq:ama}, due to the term proportional to $\lambda$, cannot be written as the gradient flow on any free energy landscape. This contrasts with the population dynamics model(s) considered later. 

Considering first a flat (i.e., one-dimensional) interface $\varphi_0$, we have 
\begin{align}\label{eq:tmu}
\mu_0[\varphi_0] &= -K\varphi_0'' +\lambda \varphi_0'^2 + f'(\varphi_0)\;,
\end{align}
while the $1/r$ coefficient of the chemical potential is given by
\begin{align}\label{eq:mu1-ama}
    \mu_1[\varphi_0] = -K(d-1)\,\varphi_0'\,.
\end{align}
Hence the linear operator associated with $\mu_0$ is given by
\begin{align}
\mathcal{L}u &= -Ku'' + 2\lambda \varphi_0' u' + u \partial_{\varphi_0}^2 f,\label{eq:mathcalL}
\end{align}
which holds for any function $u(x)$. We next explicitly find the kernel of $\mathcal{L}^\dagger$ where
\begin{align}\label{eq:Ldagama}
\mathcal{L}^\dagger \psi' = -K\psi''' - 2\lambda ( \varphi_0' \psi')' + \psi' \partial_{\varphi_0}^2 f \;.
\end{align}
We are interested in finding $\psi'\in \mathrm{Ker}(\mathcal{L}^\dagger)$ when $v_0$ is small (or, equivalently, the critical radius is large). 

As noted in Ref.~\cite{prl}, there exist pairs of $(\lambda_*,h_*)$ where $v_0=0$. The resulting curve on the $(\lambda,h)$ plane --- which can be equally viewed as either the function $\lambda_*(h)$ or  $h_*(\lambda)$ --- represents the locus where the homogeneous phases $\phi_{1,2}$ exchange global stability under variation of either the field $h$ (a familiar scenario) or the activity parameter $\lambda$. 

Translational invariance implies that
\begin{align}\label{eq:muprime}
\mathcal{L}_{\lambda_*,h_*}\varphi_0'=0
\end{align}
where $\mathcal{L}_{\lambda_*,h_*}$ is the operator $\mathcal{L}$ evaluated at $\lambda=\lambda_*, h=h_*$. By substituting the choice
\begin{align}   \label{eq:ker1}
\psi' = \varphi_0' e^{-2\lambda_*\varphi_0 /K } \end{align}
in Eq.~\eqref{eq:Ldagama} and using Eq.~\eqref{eq:muprime} we confirm that $\mathcal{L}^\dagger_{\lambda_*,h_*}\psi'=0$.
We now consider $\mathcal{L}$ with $(\lambda, h) = (\lambda_*+\delta\lambda, h_*+\delta h)$ and $|\delta \lambda|, |\delta h| $ small.  As we have $\mathcal{L}^\dagger\psi' = \mathcal{L}^\dagger_{\lambda_*,h_*} \psi' + \mathcal{O}( \delta \lambda,  \delta h)$, we conclude that
\begin{align}
c \varphi_0' e^{-2\lambda_* \varphi_0/K} + \mathcal{O}(\delta\lambda, \delta h)\in \mathrm{Ker}(\mathcal{L}^\dagger)\,.\label{eq:ker2}
\end{align}
Here $c$ is a constant that cancels in the following and can thus be set to unity. The result \eqref{eq:ker2} means that, to the order in small quantities required for our NNT calculation (and also for the dynamics of capillary waves in Sec.~\ref{subsec:cap_ama} below) the choice \eqref{eq:ker1}, when used in the reaction-coordinate definition $I(R)=0$ via \eqref{eq:defn}, avoids any further computation of the TRR-breaking corrections that would otherwise be required.

\subsection{Dynamics of the flat interface}\label{sec:AMA-1d-interface}
We calculate in this Section the dynamics of the one-dimensional moving interface for AMA. Using pseudo-spectral methods, we numerically solve the full deterministic dynamics of Eq.~\eqref{eq:ama} for the one-dimensional case and determine its velocity. In Fig.~\ref{fig:ama_together}(a) is displayed $v_0$ as a function of $\lambda$ and $h$. The symmetric structure of the velocity as a function of $(h,\lambda)$ is related to the symmetry $(h,\lambda,\phi) \to -(h,\lambda,\phi) $ present in AMA. 

Analytical progress can be made to find the values of $(\lambda_*,h_*)$ at which the interface is stationary in the absence of noise. Setting Eq.~\eqref{eq:tmu} equal to zero (static solution) and substituting $w_\mathrm{st}(\varphi_0)=\varphi_0'^2$ (where $w_\mathrm{st}$ is seen as a function of the density), the equation for $w_\mathrm{st}$ becomes 
\begin{align}
\frac{K}{2}\frac{dw_\mathrm{st}(\varphi_0)}{d\varphi_0} - \lambda_* w_\mathrm{st} - \partial_{\varphi_0} f_{h=h_*}=0\;,
\end{align}
and thus
\begin{align}\label{eq:w0}
w_\mathrm{st}(\varphi_0) &= \frac{1}{4\lambda_*^4}\left[ W(\varphi_0;\lambda_*) - W(\phi_2;\lambda_*) \right]e^{2 \lambda_*  (\varphi_0 -\phi_2)/K}\;,
\end{align}
with $W(\varphi_0;\lambda) = -4\lambda^3 \partial_{\varphi_0} f - 2K\lambda^2 \partial^2_{\varphi_0} f -6K^2\lambda\varphi_0-3K^3$. By fixing $h=h_*$ we find numerically $\lambda_*$ such that $w_\mathrm{st}(\phi_1)=0$.
The black solid line in Fig.~\ref{fig:ama_together}(a) is the locus of pairs $(\lambda_*,h_*)$ where the interface is static and thus $v_0=0$.

Knowing $w_\mathrm{st}$ and $\psi'$ allows us to calculate the stochastic dynamics of a moving one-dimensional interface in the regime where the deterministic drift is small. By considering $\lambda=\lambda_*+\delta\lambda$ with $\delta\lambda\ll 1$, we find Eq. \eqref{eq:1d-xdot}, with
\begin{align}\label{eq:v0g}
v_0 = \frac{\Delta g}{-\int_{\phi_2}^{\phi_1}  d\varphi_0 \sqrt{w_\mathrm{st}} e^{-2\lambda_* \varphi_0/K}}+\mathcal{O}(\delta \lambda^2)\;,
\end{align}
and  
\begin{align}
\mathcal{M}_{d=1} = \frac{-\int_{\phi_2}^{\phi_1} d\varphi \, \sqrt{w_\mathrm{st}}e^{-4\lambda_*\varphi/K}}{\left[\int_{\phi_2}^{\phi_1} d\varphi \, \sqrt{w_\mathrm{st}}e^{-2\lambda_*\varphi/K}\right]^2} + \mathcal{O}(\delta \lambda)
\end{align}
Here, $\Delta g:= g(\phi_1)-g(\phi_2)$, with $ g(\phi) =  \int^\phi d \varphi \, \partial_{\varphi } f \, e^{-2\lambda\varphi/K}$; explicitly, 
\begin{align} \label{eq:gdef}
g(\phi) = \frac{K}{8\lambda^4}W(\phi;\lambda)e^{-2\lambda \phi/K}+\frac{3K^4}{8\lambda^4}-\frac{K^2}{4 \lambda^2}+ \frac{hK}{2\lambda}\;,
\end{align}
where we have choosen an arbitrary constant entering in $g$ such  that $g\to f$ as $\lambda \to 0$. We note that construction $\Delta g =0$ for $(\lambda_*,h_*)$ and thus for $\lambda=\lambda_* + \delta \lambda$, $\Delta g \propto \delta \lambda$.
We plot the mobility (normalized by its equilibrium value) for different points on the $(\lambda_*,h_*)$ curve in Fig.~\ref{fig:ama_together}(b), showing its dependence on the activity parameter $\lambda_*$.

\subsection{Mobility and Quasipotential}\label{sec:NNT-AMA}
We now apply our NNT formalism to Active Model A. For $h<0$ and $\lambda=0$ (the equilibrium limit) nucleation is possible only from $\phi_1<0 \to \phi_2>0$ as 
$\phi_1$ has a higher free-energy density than $\phi_2$. As shown in Fig.~\ref{fig:ama_together}(a,d), activity can invert the relative stability of $\phi_{1,2}$; in that case $v_0$ (defined as the invasion velocity of $\phi_2$ into $\phi_1$) is positive for $h>0$ and $\lambda<0$. 

We consider the nucleation for arbitrary $h=h_*$ and $\lambda=\lambda_*+\delta \lambda+\mathcal{O}(\delta \lambda^2)$. This point lies close to the locus of neutral stability between $\phi_{1,2}$ (see Fig.~\ref{fig:ama_together}(a)) so that the interfacial velocity $v_0$ is $\mathcal{O}(\delta\lambda)$ and hence a small parameter. Using Eq.~\eqref{eq:rc-gen} we find the critical radius as
\begin{align}\label{eq:Rcinvg}
R_\mathrm{c}^{-1} = \frac{\Delta g}{(d-1)\sigma}\;,
\end{align}
where $\Delta g$ is defined via \eqref{eq:gdef} and the explicit expression for $\sigma$ is 
\begin{align}\label{eq:sigmaA}
    \sigma = K\int_r \psi'\varphi_0'=K\int_{\phi_1}^{\phi_2}  d\varphi_0 \sqrt{w_\mathrm{st}} e^{-2\lambda_* \varphi_0/K}+\mathcal{O}(\delta \lambda).
\end{align}
The expression in Eq.~\eqref{eq:Rcinvg} is valid only when $R_\mathrm{c}^{-1} \geq 0$; otherwise, the critical radius diverges (equivalently, $v_0$ changes sign) because the initial phase is already the stable one. In
Fig.~\ref{fig:ama_together}(e) we plot (for $d=2$) $R_\mathrm{c}(\lambda)$ for three fixed values of $h=h_*$; as expected $R_\mathrm{c}\to\infty$ when approaching $\lambda = \lambda_*(h_*)$ at which stability of the two phases is interchanged.

From Eq.~\eqref{eq:mob-gen}
it follows that 
\begin{align}\label{eq:amamob}
\mathcal{M}(R) = \frac{1}{R^{d-1}S_d} \frac{-\int_{\phi_2}^{\phi_1} d\varphi \, \sqrt{w_\mathrm{st}}e^{-4\lambda_*\varphi/K}}{\left[\int_{\phi_2}^{\phi_1} d\varphi \, \sqrt{w_\mathrm{st}}e^{-2\lambda_*\varphi/K}\right]^2}+\mathcal{O}(\delta \lambda)\;.
\end{align}
From this expression, the quasipotential $U(R)$ follows by integration of $v(R)/\mathcal{M}(R)$ where $v(R)$ represents the deterministic drift in \eqref{eq:gen_Rdot_drop}; for AMA this reads as $v(R) = R_\mathrm{c}^{-1}- R^{-1}$ with $R_\mathrm{c}$ given in Eq.~\eqref{eq:Rcinvg}.
Therefore, combining Eqs.~\eqref{eq:Rcinvg}--\eqref{eq:amamob}
we obtain the barrier height to be 
\begin{align}\label{eq:Uama}
U(R_\mathrm{c}) =  \frac{S_d R_\mathrm{c}^{d-1}\sigma\left[\int_{\phi_2}^{\phi_1} d\varphi \, \sqrt{w_\mathrm{st}}e^{-2\lambda_*\varphi/K}\right]}{-d\int_{\phi_2}^{\phi_1} d\varphi \sqrt{w_\mathrm{st}} \, e^{-4\lambda_*\varphi/K}}.
\end{align}
This can be compared with the quasipotential barrier derived from the TRR ansatz [Eq.~\eqref{eq:Uphi}], which for AMA is given by
\begin{align}\label{eq:Uvarphiama}
U_\varphi(R_\mathrm{c}) = \frac{-S_d R_\mathrm{c}^{d-1}K\int_{\phi_2}^{\phi_1} d\varphi \sqrt{w_\mathrm{st}}}{d}\;.
\end{align}
The corresponding barrier heights (in two dimensions) are shown in Fig.~\ref{fig:ama_together}(f), where there is a significant difference between $U(R_\mathrm{c})$ (solid lines) and $U_\varphi(R_\mathrm{c})$ (dotted lines). The NNT and TRR barrier calculations coincide on the approach to equilibrium ($h,\lambda\ll 1$) (gray line). This is expected because the TRR ansatz always holds in equilibrium. Moreover it allows a perturbative analysis close to this limit, which we describe next.

\subsection{Near-equilibrium perturbative solution}\label{sec:AMA-near-eq}
We now show that the dynamics of the flat interface can be analysed in a simple closed form, perturbatively in the limit where $h$ and $\lambda$ are both small. For simplicity we fix $K=1$. This allows us also to obtain the quasipotential in the same limit. 

Under these near-equilibrium conditions, $\psi' \simeq \varphi'_0$.
We are first interested in obtaining a perturbative solution for the one-dimensional interfacial shape $\varphi_0(x-X(t))$ in the absence of noise. The corresponding equation for $X(t)$ is given by  $     \varphi_0'' + \dot X\varphi_0' - \lambda \varphi_0'^2 - \partial_{\varphi_0} f(\varphi_0) = h$. 
The boundary conditions are $\varphi'(\infty) = 0$ and $\varphi(\infty) = \phi_1$.

We consider the perturbative solution $    \varphi_0(x-X) =\varphi_{0,0} + \lambda \varphi_{0,1} +  h \varphi_{0,2} + \mathcal{O}(\lambda^2,h^2,\lambda h)$, and $\dot X = \dot X_0 + \lambda \dot X_1  + h \dot X_2 + \mathcal{O}(\lambda^2,h^2,\lambda h )$.
The solution for $\varphi_{0,0}$ takes the usual form as $    \varphi_{0,0}(x) = - \tanh\left( \frac{x}{\sqrt{2}}\right)$,
where the minus sign arises from our convention concerning which phase is which.
By solving these equations, we find that the interfacial velocity coefficients are $\dot X_0 = 0$, $\dot X_1=2\lambda\sqrt{2}/5$, and $\dot X_2 = 3h/\sqrt{2}$.
At leading order in $\lambda $ and $h$ the perturbative solution of $\varphi_0(x)$ is found as
\begin{align}
    \varphi_0(x) &= -\tanh\left( \frac{x}{\sqrt{2}}\right) -  \frac{h}{2}\tanh^2\left( \frac{x}{\sqrt{2}}\right) \nonumber\\&+ \lambda  \frac{2\ln\frac{ 1 + \exp\left( x\sqrt{2}\right)}{2} -x\sqrt{2} }{10}\mathrm{sech}^2\left( \frac{x}{\sqrt{2}}\right).
\end{align}

Having found the interface profile and the drift velocities perturbatively for $\lambda,h\ll1$, we can obtain the quasi-potential in this limit. In particular, using Eq.~\eqref{eq:rc-gen}, the
critical radius (gray line in Fig.~\ref{fig:ama_together}(d)) obeys
\begin{align}\label{eq:pertrc}
R_\mathrm{c} = (d-1)\left(-\frac{2\sqrt{2}\lambda}{5} - \frac{3h}{\sqrt{2}}\right)^{-1} + \mathcal{O}(\lambda^2,h^2,\lambda h)\;,
\end{align}
and from Eq.~\eqref{eq:mob-gen}, we find that the mobility is  $\mathcal{M}(R)= 3R^{1-d}S_d^{-1} 8^{-1/2} + \mathcal{O}(\lambda^2,h^2,\lambda h)$.
Therefore, the barrier height is
\begin{align}\label{eq:Usmall}
U(R_\mathrm{c})=\frac{S_d R_\mathrm{c}^{d-1}\sqrt{8}}{3d},
\end{align}
which in $d=2$ becomes $U(R_\mathrm{c})=-20\pi(12\lambda +45h)^{-1}$~\cite{prl} and corresponds to the gray 
line in Fig.~\ref{fig:ama_together}(e).

\subsection{Alternative expressions for the interfacial drift}

As discussed in~\cite{prl} and in Sec. \ref{sec:general-theory}, the deterministic part of the dynamics of the droplet radius $R$ and of the interface location $X$ is independent of the choice of $\psi'$. This is at odds with the noise amplitude, encoded in the mobilities $\mathcal{M}$ and $\mathcal{M}_{d=1}$, as it follows directly from the definition of $\varphi$ and $\varphi_0$ given in Sec. \ref{sec:int-position}.

To illustrate this point more explicitly, we compute the drift of the flat interface, which can be expressed as
$v_0^\psi = \int_r \mu_0 \psi'/\int_r \varphi'_0\psi'$, where $\psi'$ lies in the $\mathrm{Ker}(\mathcal{L}^\dagger)$, or as $v_0^{\varphi_0} = \int_r \mu_0 \varphi_0'/\int_r \varphi_0'^2$. Even though these two expressions look different they give exactly the same value, as shown in Fig.~\ref{fig:ama_together}(c). 

We can also confirm this analytically at the perturbative level. By considering a slowly moving interface [$\lambda = \lambda_* + \delta \lambda + \mathcal{O}(\delta \lambda^2)$] and introducing $w=\varphi_0'^2$ and setting $K=1$, we find that
\begin{align}
\frac{1}{2}\frac{dw}{d\varphi_0} - \lambda w - \partial_{\varphi_0} f=v_0\sqrt{w}\;,
\end{align}
where $\varphi_0$ is a moving profile. Since $v_0$ is small, we find perturbatively that the velocity is given by
\begin{align}\label{v0-blah}
    v_0 = \frac{w_\mathrm{st}(\phi_1,\lambda)}{-2 e^{2\lambda_* \phi_1}\int_{\phi_2}^{\phi_1} d\varphi\sqrt{w_\mathrm{st}(\varphi)} e^{-2\lambda_* \varphi}}+\mathcal{O}(\delta \lambda^2)\,.
\end{align}
Using Eq. \eqref{eq:w0}, it is straightforward to see that Eq. \eqref{v0-blah} and ~\eqref{eq:v0g} exactly coincide.

\subsection{Capillary fluctuations for AMA}\label{subsec:cap_ama}
In deriving NNT, we have assumed that the fluctuations around the relaxation trajectory (which contribute to $\epsilon$ or $\epsilon_\mathrm{A}$) remain stable. While we have focused mainly on the radial density profile, a necessary condition for stability of the instanton is that transverse (capillary-wave) deformation of the interface are stable. Care is then needed, because it is known in other systems that activity can lead to capillary-wave instability~\cite{PhysRevLett.127.068001,Cates2025May}. 

In this Section we show that in AMA, the capillary wave tension $\sigma_\mathrm{cw}$ remains positive. This implies that fluctuations away from a spherical droplet shape are, within the small noise regime relevant for NNT, of order $\mathcal{O}(\sqrt{T/\sigma_\mathrm{cw}})$ and can safely be ignored. 

Because NNT is derived in the limit of large $R$ and $R_\mathrm{c}$, we address here the long-wavelength dynamics of capillary waves when the noiseless system has a stationary flat interface, {\em i.e.}, on the $v_0 = 0$ curve $(\lambda,h)=(\lambda_*,h_*)$. As usual we work in the regime of weak noise,  $T\ll 1$.
We assume that the interface has no overhangs, and lies (on average) perpendicular to the $y$ direction. We denote by $\p{r}$ the $d-1$ transverse coordinate(s), and the interfacial height by $\hat h(\p{r}, t)$. Following the methods of Sec.~\ref{sec:int-position}, the interfacial height is defined as the solution to 
\begin{align}\label{eq:capdefn}
    0 = \int_y \, \epsilon_\mathrm{cw}(\p{x},\hat h(\p{r},t),t)  \psi'(y-\hat h(\p{r},t))\,,
\end{align}
where $\epsilon_\mathrm{cw}(\p{x},\hat h(\p{r},t),t)= \phi(\p{x},t)-\varphi_0(y-\hat h(\p{r},t))$, $\int_y$ is the integral across the interface, and $\varphi_0$ is the profile of the noiseless flat interface. As was the case in NNT, we assume $\epsilon_\mathrm{cw}$ to be small and show this assumption to be self-consistent {\em a posteriori}. Likewise we temporarily leave $\psi'(y)$ unspecified. 

By taking the time derivative of Eq.~\eqref{eq:capdefn}, we find
\begin{align}\label{eq:chi_full}
    \partial_t \hat{h} =   -\frac{\int_y\dot \phi \, \psi'}{\int_y \psi' \varphi_0' - \int_y \psi'' \epsilon_\mathrm{cw}}\;.
\end{align}
Let us consider the numerator of this expression. Expanding for small $\epsilon_\mathrm{cw}$ and small $h$ we find 
\begin{align}\label{eq:ama-cw-int}
-\int_y\dot \phi \, \psi' 
&=
\int_u \psi'(u) \mu[\varphi_0(u)]
+ K(\nabla_{\p{r}}^2 \hat h)\int_u \psi'(u)\varphi_0'(u)\nonumber\\
&-\nabla_{\p{r}}^2 \int_y \psi'(y) \epsilon_\mathrm{cw}
+\int_y \psi'(y) \mathcal{L}\epsilon_\mathrm{cw} \\
&-\sqrt{2T}\int_y \psi'(y-\hat h) \xi
+\mathcal{O}(\epsilon_\mathrm{cw}^2,\nabla_{\p{r}}^2\hat h^2,\nabla_{\p{r}}^2\epsilon_\mathrm{cw} \hat h)
\;,\nonumber
\end{align}
where in the first term we have changed variable to $u=y-\hat h(\p{r},t)$ and $\mathcal{L}$ is the linear operator in Eq.~\eqref{eq:mathcalL}. The first term in the right hand side of \eqref{eq:ama-cw-int} vanishes because the interface is stationary when $T=0$. Moreover, the second line of \eqref{eq:ama-cw-int} vanishes at the order considered because of Eq.~\eqref{eq:capdefn}, so long as we choose $\psi'\in\mathrm{Ker}(\mathcal{L}^\dagger)$ as usual. 

Hence with this choice 
\begin{align}\label{eq:ama-cw-int-2}
-\int_y\dot \phi \, \psi' 
&=
K(\nabla_{\p{r}}^2 h)\int_u \psi'(u)\varphi_0'(u)\\
&-\sqrt{2T}\int_y \psi'(y-\hat h) \xi
+\mathcal{O}(\epsilon_\mathrm{cw}^2,\nabla_{\p{r}}^2\hat h^2,\nabla_{\p{r}}^2\epsilon_\mathrm{cw} \hat h)
\,.\nonumber
\end{align}
Injecting \eqref{eq:ama-cw-int-2} into \eqref{eq:chi_full} and expanding for small $\epsilon_\mathrm{cw}$ we thus conclude that 
\begin{align}\label{eq:EW}
    \partial_t \hat h =  \frac{\sigma_\mathrm{cw}}{\int_y \psi' \varphi_0'} 
    \nabla_\p{r}^2\hat h + \sqrt{ \frac{2T \int_y \psi'^2}{(\int_y \psi'\varphi_0')^2}}\;\eta(\p{r},t)\,,
\end{align}
where $\langle\eta(\p{r},t) \eta(\p{r}',t')\rangle = \delta(t-t') \delta(\p{r}-\p{r}')$ and 
$\sigma_\mathrm{cw} = \sigma$, where $\sigma $ is given in eq. \eqref{eq:sigmaA}. 
In obtaining \eqref{eq:EW}, we have neglected terms of order $\mathcal{O}(\epsilon_\mathrm{cw}^2,(\nabla_{\p{r}} h)^2,\nabla_{\p{r}}^2 \epsilon_\mathrm{cw} \hat h, \epsilon_\mathrm{cw}\sqrt{T})$. In any case, since $\psi$ and $\varphi_0$ are both monotone decreasing, this tension is always positive.

It remains to be shown that $\epsilon_\mathrm{cw}$ is small as assumed. Computing $\partial_t \epsilon_\mathrm{cw}$ we find  
\begin{align}
(\partial_t +\mathcal{L}-\nabla_{\p{r}}^2)\epsilon_\mathrm{cw}
=
\varphi_0' \partial_t \hat{h} + \sqrt{2T} \xi
+\mathcal{O}(...)\,.
\end{align}
Here $\mathcal{O}(...)= \mathcal{O}(\epsilon_\mathrm{cw}^2,(\nabla_{\p{r}} h)^2, \nabla_{\p{r}}^2\epsilon_\mathrm{cw} \hat h, \epsilon_\mathrm{cw}\sqrt{T})$ and $\xi$ is the noise entering in Eq.~\eqref{eq:genmodel} for the density field. 
Neglecting for the moment $\nabla_{\p{r}}^2\epsilon_\mathrm{cw}$, we find that $\epsilon_\mathrm{cw}\sim \mathcal{O}(\nabla_{\p{r}}^2 h, \sqrt{T} )$. It then follows that terms in $\mathcal{O}(...)$ are negligible. The same applies to 
$\nabla_{\p{r}}^2 \epsilon_\mathrm{cw}\sim \mathcal{O}(\nabla_{\p{r}}^4 \hat{h}, \sqrt{T}\nabla_{\p{r}}^2\hat h)$. Hence, we conclude that terms neglected in deriving \eqref{eq:EW} are of order $\mathcal{O}((\nabla_{\p{r}} h)^2, \sqrt{T}\nabla_{\p{r}}^2 h, T)$ and indeed subleading in \eqref{eq:EW}. This concludes the self-consistent argument.

Eq.~\eqref{eq:EW} is the Edwards-Wilkinson equation with  noise amplitude set by activity via $\psi'$. 
It was also recently obtained with a method based on the path-integral representation of the AMA dynamics in~\cite{Sarfati2026May}. 

Extending the analysis developed in this Section to
generic non-conserved models obeying Eq.~\eqref{eq:genmodel} is straightforward but not pursued here.  Moreover, the same analysis can be used for conserved models and in known cases (AMB+ and Active Cahn-Hilliard equations) it reproduces the stochastic dynamics of capillary waves previously obtained in the literature~\cite{PhysRevLett.127.068001,besse2023interface,caballero2025interface,burekovic2026active}.

A difference between the non-conserved and conserved cases is that, in the latter, the same results are obtained for {\em any} choice of $\psi'$, since the conservation law ensures that to the required order, $\epsilon_\mathrm{cw}=0$. In contrast, the term containing $\mathcal{L}\epsilon_{\mathrm{cw}}$ in Eq.~\eqref{eq:ama-cw-int} does not vanish if $\psi'\notin \mathrm{Ker}(\mathcal{L}^\dagger)$, and it modifies the noise variance. Mirroring precisely what we found in comparing NNT with the TRR ansatz, Eq.~\eqref{eq:EW}  correctly describes the stochastic dynamics of capillary waves solely when $\psi'\in \mathrm{Ker}(\mathcal{L}^\dagger)$. For AMA we recall from Sec.~\ref{sec:AMA-ker} that if so, $\psi'=\varphi_0' e^{-2\lambda_* \varphi_0}$. 

\section{Population dynamics}\label{sec:pd}
In spatially extended population dynamics models, 
a spatiotemporal field $\phi(\p{r},t)$ can represent (for example) either the local density of a single species, or the local fraction of a binary subtype within a spatially uniform population density -- the latter representing a fitter mutant, an invading species, or the like.
The field-level dynamics is typically described by a reaction diffusion equation \cite{cantrellSpatialEcologyReactionDiffusion2004, garcia2012noise,korolevGeneticDemixingEvolution2010}     
\begin{align}\label{eq:PDgen}
    \partial_t \phi = -\partial_\phi f_\mathrm{PD} (\phi) + K \nabla^2 \phi + \sqrt{2 T D(\phi)}\,\xi,
\end{align}
where $-\partial_\phi f_\mathrm{PD} (\phi)$ is the reaction term, $K$ the diffusion constant, and $\xi(\p{r},t)$ a unit white Gaussian noise term. The factor $T$ plays a similar role to the temperature in equilibrium models, but now is inversely related to the overall population density scale. 
\Eqref{eq:PDgen} is equivalent to the general formulation \Eqref{eq:genmodel} with
\begin{align}
    \mu[\phi] = \partial_\phi f_\mathrm{PD} (\phi) - K \nabla^2 \phi\;.
\end{align}

Note that unlike AMA, this $\mu[\phi] = \delta \mathcal{F}/\delta\phi$ for some $\mathcal{F}$ so the noiseless dynamics is always a gradient flow. 
However, since both the reaction part $-\partial_\phi f_\mathrm{PD} (\phi)$ and the noise $D(\phi)$ depend on the details of the reproduction dynamics, these generically do not follow a fluctuation dissipation relation; doing so would require not $\mu[\phi]$ but rather $\mu[\phi]/D(\phi)$ to be a functional derivative.
Hence,  \Eqref{eq:PDgen} is out of equilibrium and the tools of NNT are required for describing nucleation.

\begin{figure}[t]
    \centering
\includegraphics[width=\columnwidth]{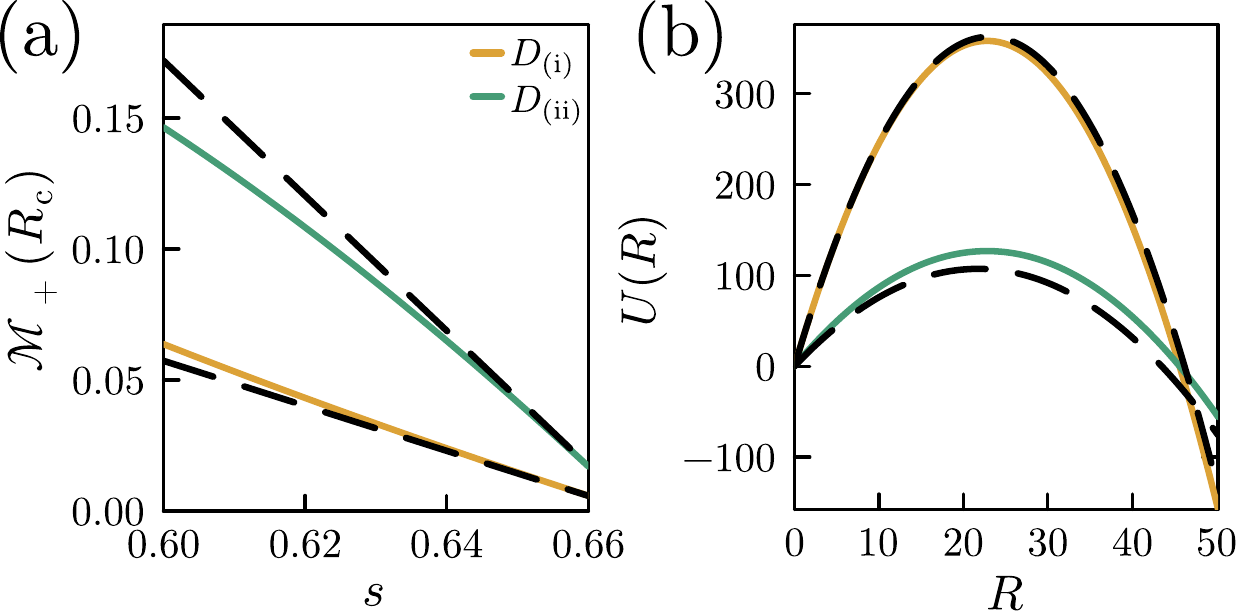}
    \caption{Mobility (a) and quasi-potential (b) for the population dynamics model variants. The colored lines correspond to the solutions from the full profile $\varphi_0$ (from evaluating numerically Eq.~(\ref{eq:mob-gen}) in panel (a) and Eq.~(\ref{eq:URgen}) in panel (b)) and the dashed lines from perturbation theory in small $s-2/3$. To obtain the continuum lines, the profile $\varphi_0$ was found by numerically solving \Eqref{eq:popModel} with a simple explicit Euler scheme for $T=0$.
    The quasi-potentials are evaluated for $s=0.65$ and we considered $D_0=10^{-3}$ throughout.} 
    \label{fig:fig-pop-dyn-pert}
\end{figure}

In the following, we will treat two examples that differ only in the form of $D(\phi)$. In the first (case (i)), a uniform population density has two subtypes, A and B, and the field $\phi$ describes the local fraction of A (say).
The rate of change of this fraction is given by \cite{korolevGeneticDemixingEvolution2010}
\begin{align}\label{eq:popModel}
    \partial_t \phi = a(\phi) \phi(1-\phi) + K\nabla^2\phi + \sqrt{2T D(\phi) } \zeta, 
\end{align}
where $\phi(1-\phi)$ gives the probability of proximity between unlike individuals and $a(\phi)$ quantifies the advantage of one subtype over the other. For concreteness we choose a specific form derived from gene-drive dynamics, namely  \begin{align}\label{eq:aphi}
a(\phi)=s(\phi-\tfrac{2s-1}{s})\,, 
\end{align} 
with an advantage parameter $s\in[\tfrac{1}{2}, \tfrac{2}{3}]$~\cite{tanakaSpatialGeneDrives2017}. Note that the drift term is symmetric (about $\phi=1/2$) for the choice $s = 2/3$.

Alongside this we choose  $D(\phi)= D_{(\mathrm{i})}(\phi)$ with
\begin{align}\label{eq:Dform}
 D_{(\mathrm{i})}(\phi):=|\phi(1-\phi)|+D_0  \,. 
\end{align}
Here the first term can be derived using `stepping-stone' models \cite{korolevGeneticDemixingEvolution2010}
 \footnote{Alternatively this contribution to $D$ can be viewed as arising from additional, equiprobable random processes AB$\rightarrow$AA and AB$\rightarrow$BB.}. This term vanishes at $\phi =0,1$ which, given also the form of the drift term, limits the dynamics to the range $\phi\in[0,1]$ everywhere. Therefore the modulus signs in \eqref{eq:Dform} would be redundant, but for the additional constant term $D_0$ which allows escape from the otherwise absorbing states at $\phi=0,1$ (for instance, this could represent A$\leftrightarrow$B by random mutation). Without such escape, there would be no nucleation problem to discuss (our theory applies to the case $D_0>0$). In this case the modulus signs in \eqref{eq:Dform} ensure that $D$ remains positive everywhere. This is required for that problem to remain globally well-posed, although since $D$ is anyway positive along the instanton itself, there is no effect on the resulting action.

As a second variant, we consider a Schlögl model~\cite{Schuttler2024Aug, schloglChemicalReactionModels1972}, comprising a birth and death process with respective rates $\lambda^+(\phi)=(3s-1)\phi^2$ and $\lambda^-(\phi)=s\phi(\phi^2 + \tfrac{2s-1}{s})$. 
The dynamics of the resulting population density $\phi(\p{x},t)$
can again be written in the form of Eq.~(\ref{eq:popModel}), with $a(\phi)= \lambda^+- \lambda^- $ obeying \eqref{eq:aphi} as before. However, now $D(\phi) = D_{(\mathrm{ii})}(\phi)$ where~\cite{Schuttler2024Aug, schloglChemicalReactionModels1972}
\begin{align}
D_{(\mathrm{ii})}(\phi)-D_0= \lambda^++\lambda^-=s\phi(1+\phi)(\phi+\tfrac{2s-1}{s})\,.
\end{align}
Here, small $D_0$ again allows escape from the initial $\phi=0$ state.

\subsection{Quasipotentials for population dynamics}

We now apply NNT to both model variants to compute the quasipotential and the interface mobility.

In both cases the deterministic drift is $-\partial_\phi f_\mathrm{PD} (\phi)=a(\phi)$.
To follow our general approach, we can identify  
\begin{align}
    \mu_0[\varphi_0] = -K\varphi_0'' +\partial_{\varphi_0} f_\mathrm{PD}\\
    \Lmu u = -Ku'' + u \partial_{\varphi_0}^2 f_\mathrm{PD}\;.
\end{align}
In this case $\Lmu=\Lmu^\dagger$ and $\psi'=\varphi_0'$ from translational invariance.
The projected dynamics using $\psi'=\varphi_0'$ then reads as
\begin{align}
    \dot R =  \frac{\int_r\varphi_0' \partial_{\varphi_0} f_\mathrm{PD} }{ \int_r \varphi_0'^2}
    - \frac{(d-1)K}{R}  +  \sqrt{\frac{2 T \int_r D(\varphi_0)\varphi_0'^2 }{S_d R^{d-1}(\int_r \varphi_0'^2 )^2} } \, \eta,
\end{align}
where we deduce that $R_\mathrm{c}=\frac{(d-1)K \int \varphi_0'^2 dx}{\Delta f_\mathrm{PD}}$ with $\Delta f_\mathrm{PD}=f_\mathrm{PD}(1)-f_\mathrm{PD}(0) $. The quasipotential is simply given by Eq.~\eqref{eq:URgen}, for $\psi'=\varphi'_0$. 

\begin{figure*}[t]
\centering
\includegraphics[width=\textwidth]{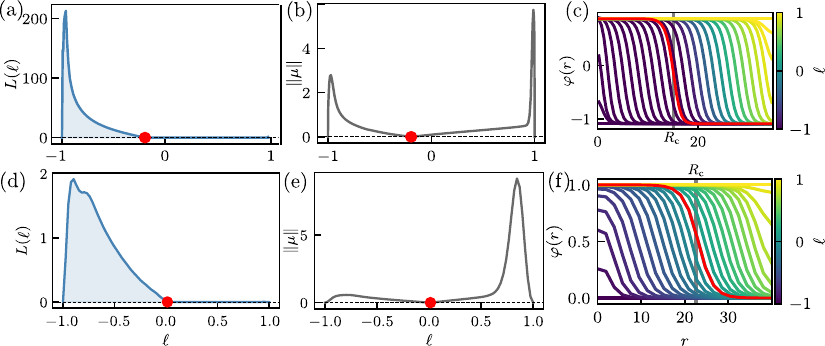}
\caption{Minimum action path for Active Model A (top row) and  Population dynamics model with $D = D_{(i)}(\phi)$(bottom row) from a Ritz method based on ~\cite{PhysRevResearch.2.033208}. (a, d) Lagrangian $L(t)$ along the minimum action path. The red dots mark the saddles, where $L(t)$ first vanishes. (b, e) Absolute value of the radially weighted drift $||\mu||$ (see Eq.~\ref{eq:radial_drift}) along the minimal action path, where the zero drift corresponds to the saddle.  (c, f) Radial order parameter field $\varphi(r)$ along minimal action path. The red lines correspond to the respective saddles where the drift vanishes. The vertical lines correspond to the extracted critical radii $R_\mathrm{c}$. Parameters for AMA are $\lambda=-1.033$, $h=0.2$ and for the population dynamics examples are $s=0.65$ and $D_0=10^{-2}$. Numerical simulations are in $d=2$ using a cell-centered radial discretizations of $\Delta r=0.5$ and a Chebyshev–Lobatto discretizations of the geometric path. For Active Model A we set an outer radius $R_{\rm out}\approx2R_{\rm c}$, chose $N_\ell=41-81$ path nodes, and evaluated the geometric action on a five-times refined interpolated grid. For the population-dynamics model we used $N_\ell=20–30$ path nodes and $N_r=16–24$ radial grid points.} 
\label{fig:RitzInsanton}
\end{figure*}

To gain analytical insight into these expressions, we look for perturbative solutions around the symmetric case $s=2/3$, addressing first a flat interface at position $X(t)$.
Imposing the $\phi(x,t) = \varphi_0(x-X(t))$ we find that $    K \varphi_0'' + \dot X \varphi_0' + \varphi_0(1-\varphi_0)(s \varphi_0 +1 - 2s)=0,$
with the condition that $\varphi_0(0)=1/2$.
The perturbative solution of the profile $\varphi_0$ and the velocity $\dot{X}$ are given by
\begin{align}\label{eq:varphi0-int-dfqeq}
    \varphi_0 &= \frac{1}{2}\left[ 1 + \mathrm{tanh}\left(\frac{x}{2\sqrt{3K}} \right)\right] + \left(\frac{2}{3}-s \right) \frac{\sqrt{\frac{3}{K}}x}{8+8 \mathrm{cosh}(\frac{x}{\sqrt{3K}})}\nonumber\\ &+ \mathcal{O}\left(\frac{2}{3}-s\right)^2\;,\\
    \dot X &= \frac{3\sqrt{3K}}{2} + \mathcal{O}\left(\frac{2}{3}-s\right)^2\;.
    \end{align} 
These corrections can be used to evaluate the quasi-potential in Eq.~(\ref{eq:URgen}) perturbatively for both (i) the gene-drive-inspired model and (ii) its Schlögel variant. 
With $K=1$ and $d=2$, the barrier heights are respectively found as
\begin{align}\label{eq:PD-U12-c}
    U_\mathrm{(i)}(R_\mathrm{c}) &= \frac{10 \pi }{\sqrt{3}(1+D_0) (2-3 s)}+ \mathcal{O}(\tfrac{2}{3}-s)\;,\\
    U_\mathrm{(ii)}(R_\mathrm{c})&=\frac{5 \pi  (-8 + 10D_0 +21 s)^2}{2 \sqrt{3}(3 + 5 D0)^3 (2-3 s)}+ \mathcal{O}(\tfrac{2}{3}-s)\;.
\end{align}
We have also solved numerically for the full interface profile $\varphi_0$. Both numerical and perturbative results are reported in Fig.~\ref{fig:fig-pop-dyn-pert} for the different cases of $D(\phi)$, showing a good agreement. It should not be surprising that the mobilities $\mathcal{M}$ are quite different for the two different noise choices, nor that the resulting quasipotential barriers are also different. 

As previously mentioned, the above calculations require $D_0>0$ to allow escape from the initial state. One might imagine for this reason that the barrier height $U(R_\mathrm{c})$ should diverge smoothly as $D_0\to 0$. However this is not the case as can be seen directly from Eq.~\eqref{eq:URgen}. This is because NNT already assumes (like CNT) that the rate-limiting step is crossing the quasipotential barrier at $R = R_\mathrm{c}$, not the formation of an initial very small droplet. The action for the latter can be estimated as $\xi^d \Delta /D_0$ with $\xi$ the interfacial width and $\Delta$ a local free energy density set by $f_\mathrm{PD}$ and $K$. Therefore NNT is valid when $D_0\gg \xi^d \Delta / U(R_\mathrm{c})$. For any finite $D_0$ this holds in the usual CNT/NNT limit of large 
$U(R_\mathrm{c})$ and Eq.~\eqref{eq:URgen} is the result of taking that limit first.


\begin{figure}[t]
    \centering
\includegraphics[width=1.0\columnwidth]{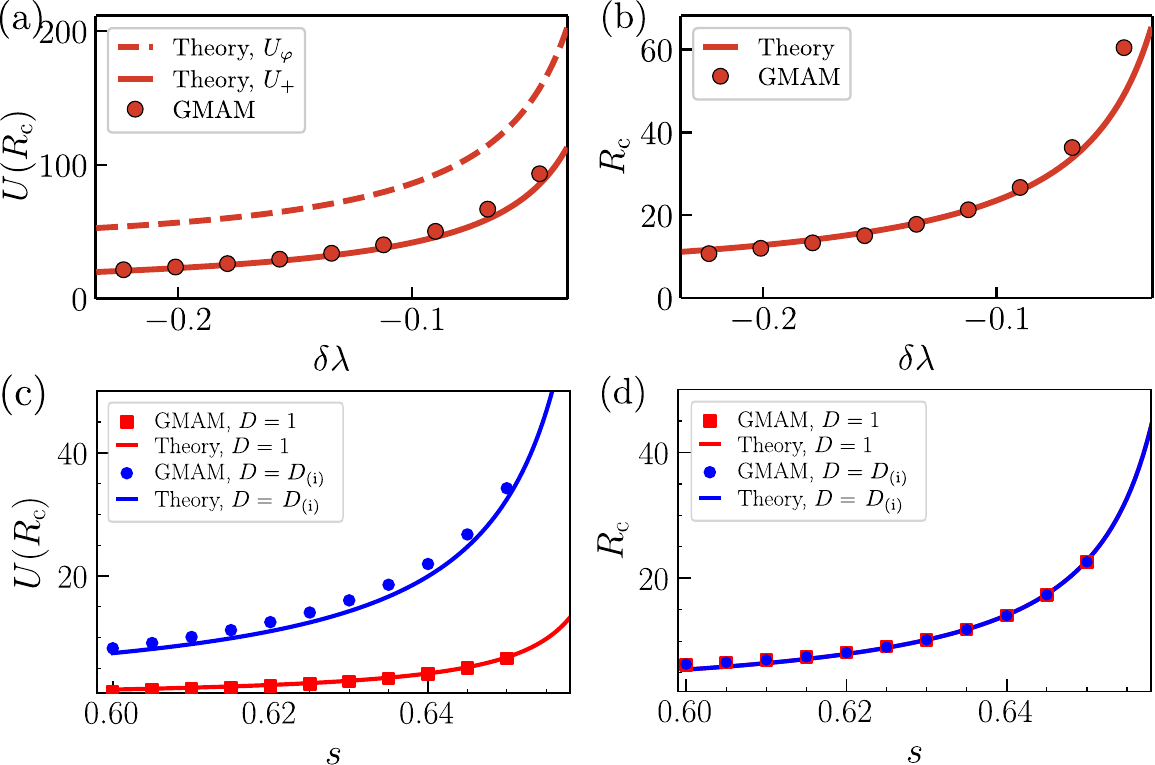}
    \caption{(a)-(b) Quasipotential barrier heights and critical radii obtained by numerical action minimization for AMA (dots) and the corresponding NNT prediction (solid line). The dashed line shows the prediction from the TRR ansatz. Data is shown for fixed $h=0.2$ and varying $\lambda =\lambda_*(h)+\delta\lambda$ where $\lambda_*(0.2)=-0.8768$. (c)-(d) Results on population dynamics, the solid line shows our NNT predictions, and the red squares are the results from the numerical minimization. Results for the equilibrium case, $D(\phi)=1$, and for $D = D_{(i)}(\phi)$, are both shown. Numerical parameters as in Fig.~\ref{fig:RitzInsanton} and $D_0=10^{-2}$. 
    }
    \label{fig:fig-pop-gmam}
\end{figure}

\subsection{Capillary fluctuations for population dynamics}\label{sec:pop-dyn-cw}
As in Sec.~\ref{subsec:cap_ama}, we can derive the dynamics of capillary waves. Since the deterministic dynamics are equilibrium-like, $\varphi_0'\in \mathrm{Ker}(\mathcal{L}^\dagger)$. This follows from translational symmetry and the fact that 
$\mathcal{L}$ is self-adjoint. 
The resulting Edwards-Wilkinson equation is then found by substituting $\varphi_0$ ({\em e.g.}, as found perturbatively in \eqref{eq:varphi0-int-dfqeq}) for $\psi$ in Eq.~\eqref{eq:EW}. The resulting capillary wave tension is  $\sigma_\mathrm{cw}=K\int_y (\varphi_0')^2$.

\section{Numerical Validation: Geometrical Minimum Action Method}\label{sec:num}

To benchmark our analytical NNT predictions, we numerically minimize the Freidlin--Wentzell action for both AMA and the population-dynamics examples.
The field dynamics are governed by Eq.~\eqref{eq:genmodel}
and the corresponding Freidlin--Wentzell action is given by Eq.~\eqref{eq:FWgen}.
Minimum action paths (MAPs) between the metastable state $\phi_1$ and the stable state $\phi_2$
are computed using the geometric minimum action method (gMAM)~\cite{heymann2008,
Grafke2019Jun}, which avoids the need for infinite-time integration.
In addition, we employ a Ritz method analogous to Ref.~\cite{PhysRevResearch.2.033208}, with finite-difference
discretization in space and a Chebyshev basis in time.

We invoke rotational symmetry to impose that the instanton density field remains a function of the radial coordinate only, $\phi(\p{r},t) = \phi(r, t)$.
The field is then discretized on a cell-centered radial grid of $N_r$ points over
$r \in (0, R_\mathrm{out}]$ with uniform spacing $\Delta r$.
Denoting the total drift at lattice site $i$ as $\mu_i=\mu[\phi_i]$,
the discretized action reads (in $d=2$)
\begin{align}
\mathcal{A} = \frac{1}{4T}\int dt \,L(t)= \frac{2\pi}{4T} \int dt \sum_i r_i\, \Delta r\,
\frac{\bigl(\partial_t \phi_i + \mu_i\bigr)^2}{D(\phi_i)}\;,
\end{align}
where we have performed the angular integral, and $L(t)$ is a discrete Lagrangian.
We re-parametrize time by the path arc length $\ell$, writing
$\partial_t \phi_i = \Lambda(\ell)\, \partial_\ell \phi_i$ with $\Lambda = d\ell/dt$, giving
\begin{align}
\mathcal{A} = \frac{\pi}{2T} \int \frac{d\ell}{\Lambda} \sum_i r_i\, \Delta r\,
\frac{\bigl(\Lambda\, \partial_\ell \phi_i + \mu_i\bigr)^2}{D(\phi_i)}\;.
\end{align}
Minimizing the integrand pointwise over $\Lambda(\ell)$ at each $\ell$ independently,
immediately yields $\Lambda(\ell) = \sqrt{{c(\ell)}/{b(\ell)}}$ with
$b(\ell) = \sum_i \frac{r_i\,\Delta r\,(\partial_\ell \phi_i)^2}{D(\phi_i)}$ and     $c(\ell) = \sum_i \frac{r_i\,\Delta r\, \mu_i^2}{D(\phi_i)}$.
Substituting back gives the geometric action
\begin{align}
A_g &= \frac{\pi}{T} \int d\ell \left[
\sqrt{\left(\sum_i \frac{r_i\,\Delta r\,(\partial_\ell\phi_i)^2}{D(\phi_i)}\right)
\!\left(\sum_i \frac{r_i\,\Delta r\, \mu_i^2}{D(\phi_i)}\right)}\right. \nonumber\\
&\left. + \sum_i \frac{r_i\,\Delta r\, \mu_i\,\partial_\ell\phi_i}{D(\phi_i)}
\right].
\label{eq:Ag}
\end{align}

\begin{figure*}[t]
\centering
\includegraphics[width=0.9\textwidth]{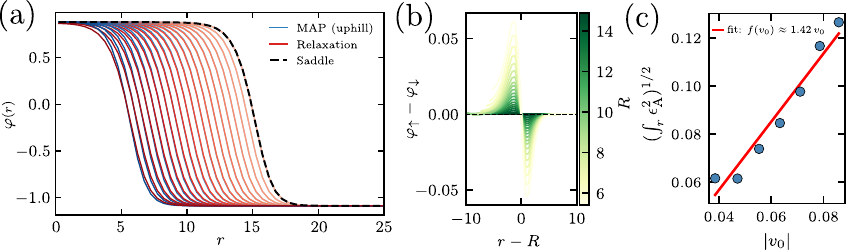}
\caption{Quantification of deviation between instanton and reversed relaxation dynamics for Active Model A. (a) Scalar order parameter profiles $\varphi(r)$ along instanton (red) obtained from numerical minimization of the geometric action and from numerical solution of \Eqref{eq:ama} starting from the saddle (blue). (b) Difference between the order parameter along the instanton and the TRR path as a function of the distance from the interface. (c) Averaging the difference between the interfacial profile along the instanton and along the relaxation trajectory over the full instanton shows that $({\int_r \epsilon_\mathrm{A}^2})^{1/2}\sim v_0$ as expected from NNT framework. The fit (red solid line) highlights the expected linear scaling with $v_0$. Each point in (c) corresponds to a different value $\delta \lambda$ as specified in Fig.~\ref{fig:fig-pop-gmam}(a)--(b). Throughout: Order parameter evolutions are obtained with $h=0.2$ and $\lambda=-1.0333$. The radial grid uses cell-centered spacing $\Delta r = 0.5$, with the outer boundary $R_\mathrm{out}$ set to $2R_\mathrm{c}$. The path was discretized using $N_s = 41$ Chebyshev--Lobatto nodes, with the geometric action evaluated on a $5\times$ finer interpolated grid ($N_\mathrm{int} = 205$ nodes).}
\label{fig:fig-epsilon}
\end{figure*}

The arc-length coordinate $\ell \in [-1, 1]$ is discretized on $N_\ell$
Chebyshev--Lobatto nodes, with the path endpoints fixed at the two equilibrium states.
The radial Laplacian $\nabla^2\phi = \phi''(r) + (d-1)\phi'(r)/r$, needed to
evaluate the drift term, is approximated by second-order finite differences.
A flux-balanced stencil at the first interior node enforces the ghost-cell
Neumann condition $\partial_r\phi\big|_{r=0} = 0$, and a zero-flux condition is
imposed at the outer boundary $r = R_\mathrm{out}$. For the system size $R_\mathrm{out}$, it is convenient to use a multiple of the analytical prediction for the critical radius, which minimizes finite size effects without recourse to very large system sizes. 

To avoid reparameterization artifacts, the action integrand is evaluated on an
oversampled grid of $N_\mathrm{int}=n_\mathrm{int}N_\ell$ Chebyshev--Lobatto points via barycentric interpolation which is also used to evaluate the derivatives along the path  $\partial_\ell\phi$. 
The $N_r(N_\ell - 2)$ interior degrees of freedom are optimized with an
L-BFGS procedure~\cite{lbfgs} implemented via
\texttt{Optim.jl}~\cite{mogensen2018optim} and \texttt{Optimizers.jl} in Julia~\cite{bezanson2017julia}. 
The result of this procedure is shown in Fig.~\ref{fig:RitzInsanton}, with the optimized Lagrangian for AMA in panel (a) and for the population dynamics model (noise variant (i)) in panel (d); the absolute value of the drift respectively in panels (b,e); and the instanton path in the order parameter field in panels (c,f).

The saddle configuration (critical nucleus) is identified as the path point at which
the weighted norm of the total drift force 
\begin{align}\label{eq:radial_drift}
\|\mu\| = \sqrt{\sum_i r_i\,\Delta r\, \mu_i^2}\,
\end{align}
is minimized along the path, see Fig.~\ref{fig:RitzInsanton} (b) and (e).
Note that this minimum coincides with the point where the Lagrangian vanishes as depicted in Fig.~\ref{fig:RitzInsanton} (a) and (d).
The identified saddle is also indicated in the order parameter fields in red (c) and (f).
The critical radius $R_\mathrm{c}$ is then extracted by linearly interpolating the saddle
profile through the midpoint value
$\phi_\mathrm{mid} = (\phi_1 + \phi_2)/2$.

For the population dynamics model, we first minimize the action with $D(\phi)=1$, where the underlying equation has an equilibrium structure.
We then use the extracted instanton as an initial guess for the nonequilibrium problem, focusing on the case of $D(\phi) = D_{(i)} (\phi)$ obeying \eqref{eq:Dform}.

As shown in Fig.~\ref{fig:fig-pop-gmam}, the critical radii and the quasipotentials calculated using the numerical action minimization are in good agreement with our NNT theory for AMA (a,b), and the population dynamics model, (c,d).
The predicted critical radii in Fig.~\ref{fig:fig-pop-gmam}(b,d) are likewise matching well, although these are not a direct test of NNT since they are follow from the deterministic dynamics alone.

We can also use the result of the minimization to quantify $\epsilon_\mathrm{A}$ as the difference between the instanton and the time-reversed relaxational path. Restricting attention to AMA, the interfacial profiles are compared in Fig.~\ref{fig:fig-epsilon}(a), and their difference $\epsilon_\mathrm{A}(r-R)$ quantified in Fig.~\ref{fig:fig-epsilon}(b). We also show in Fig.~\ref{fig:fig-epsilon}(c) that this difference scales approximately linearly with the velocity as predicted analytically in Sec.~\ref{sec:action-non-trs} above.

\section{NNT for Active Model B+}\label{sec:AMB+}

We now present NNT for Active Model B+ (AMB+), the minimal model describing active phase-separating systems~\cite{Cates2025May}. This will provide an independent derivation of Eq.~\eqref{eq:Rdot_intro} in this case, retrieving the results for the mobility and quasipotential found by other means in~\cite{catesClassicalNucleationTheory2023}. We shall see that the existence of a conserved order parameter causes the TRR ansatz, while not exactly true, to be valid to the required order for nucleation theory. This contrasts with the non-conserved cases considered above.
In Sec.~\ref{sub:AMB+stoc} we obtain Eq.~\eqref{eq:Rdot_intro} via the stochastic route. In Sec. \ref{sub:AMB+TRR} we then show that the same result can be obtained via the TRR ansatz from the action approach. 

\subsection{Stochastic route for AMB+}\label{sub:AMB+stoc}
 We focus on the $d=3$ case, but the same calculation applies for $d=2$ (up to logarithmic corrections to the mobility). AMB+ is defined by
\begin{align}
    \dot \phi = -\nabla \cdot (\p{J} + \sqrt{2T}\boldsymbol{\xi})
\end{align}
with $\p{J} = -\nabla[\frac{\delta \mathcal{F}}{\delta \phi}+\lambda (\nabla \phi)^2 + \zeta \nabla\phi \nabla^2\phi]$ and $\langle \xi_{i} \xi_{j}\rangle=\delta_{ij}\delta(t-t')\delta(\p{x}-\p{x}')$, $\delta \mathcal{F}/\delta \phi =  f'(\phi)-K\nabla^2\phi$ and  $f(\phi)$ is a double-well potential. In the following, we will consider $K=1$.
Using Helmholtz decomposition we decompose the part of the current proportional to $\zeta$ as $-\nabla \mu_\zeta + \nabla \land \p{A}$. 
Here $\mu_\zeta$ takes the non-local form~\cite{PhysRevX.8.031080}
\begin{align}
    \mu_\zeta = - \zeta \int_{\p{x}'} \nabla \cdot (\nabla \phi \nabla^2\phi)(\p{x}') G(\p{x},\p{x}').
\end{align}
where $G(\p{x};\p{x}')=(4\pi)^{-1}|\p{x}-\p{x}'|^{-1} $.
The remaining part $\nabla\land\p{A}$ does not contribute to $\dot\phi$, and can be ignored in the following.
We define $\mu = \frac{\delta \mathcal{F}}{\delta \phi}+\lambda (\nabla \phi)^2 + \mu_\zeta $. We assume the system to be in an initial state with global density $\phi_s = \phi_1+\delta_s$, where $\delta_s$ is the supersaturation, and we are interested in the nucleation of a spherical droplet of the phase $\phi_2$. (A similar calculation of course applies with the phases interchanged.) 

In conservative models, nucleation takes place by the nucleating droplet being locally quasi-stationary, with dynamics forced by both the supersaturation in the far field, and by the noise current. As such, we consider $\phi(\p{x}, t) =\varphi(r, R)+\epsilon(\mathbf{x}, t)$ where, now, $\varphi(r,R)$ is the density profile of a {\em stationary} droplet of radius $R$. (Note that such a solution can be found in any finite system by choosing the correct global density~\cite{PhysRevX.8.031080,Cates2025May}.) We now define the radius of the droplet $R$ such that
\begin{align}
    0 &= \int_\p{x} \bar\psi'(r-R)\epsilon(\mathbf{x}, t)\;,
\end{align}
where, for later convenience, we have now denoted the test function by $\bar{\psi}'$. Taking the time derivative and expanding for small $\epsilon$, we obtain 
\begin{align}\label{eq:AMB+interm} 
     &\dot R \left(\int_\p{x}\varphi'\bar{\psi}'-\int_\p{x} \bar \psi'' \epsilon \right) = \nonumber\\&-\int_\p{x} \bar\psi'(\nabla^2\mu[\varphi] +\nabla^2(\mathcal{L}_\varphi\epsilon)  - \sqrt{2T} \nabla \cdot \boldsymbol{\xi})
\end{align}
where $\mu[\phi] = \mu[\varphi]+\mathcal{L}_\varphi\epsilon$, and we are neglecting terms of order $\epsilon^2$. 

Now, by choosing $\nabla^2\bar\psi' = \psi' \in \mathrm{Ker}\mathcal{L}_\varphi^\dagger $ the $\epsilon$ term vanishes. (Note the extra Laplacian compared to the non-conserved case.) As also discussed in Sec.~\ref{sec:action-non-trs}, to leading order, $\mathcal{L}_\varphi = \mathcal{L}$ where $\mathcal{L}$ is the linearized operator associated with $\mu_0$. Hence, we choose $\psi' = e^{(\zeta-2\lambda)\varphi}\varphi' \in \mathrm{Ker}\mathcal{L}^\dagger$. (In the AMB+ context, $\psi$ is called the pseudo-density~\cite{Cates2025May}.)  We thus find that
\begin{align}
    \dot R \left(\int_\p{x} \psi' \nabla^{-2} \varphi' - \int_\p{x} \bar \psi'' \epsilon \right)  = \int_\p{x} \psi'[ f'(\phi_s) - \mu(\varphi)] \nonumber\\+ \sqrt{2T}\int_\p{x} \psi' \nabla^{-2} (\nabla \cdot \boldsymbol{\xi}) 
\end{align}
where we used that $\mu(r\to\infty) = f'(\phi_s)$.
Following similar steps as in Ref.~\cite{catesClassicalNucleationTheory2023} we then find that
\begin{align}
    \dot R &= \frac{f'(\phi_s)\Delta\psi-\int_r(\psi'\mu_0[\varphi])}{-R\Delta\phi\Delta \psi} + \sqrt{2T\mathcal{M}}\eta(t)\nonumber\\
    &+\frac{1}{R}\frac{\int_r\psi'\mu_1[\varphi]}{R\Delta\phi\Delta \psi} + \mathcal{O}(...),
\end{align}
where $\Delta \phi = \phi_s-\phi_2$, $\Delta \psi = \psi(\phi_s)-\psi(\phi_2)$ and $\eta(t)$ is a Gaussian white noise of unit variance. Here $\mathcal{O}(...)=\mathcal{O}(\epsilon R^{-2},\epsilon v_0R^{-1},\epsilon R^{-3/2} \sqrt{T},\epsilon^2,R^{-3},R^{-2}v_0)$; for $d=3$, $\mu_1$ is defined as $    \mu_1[\varphi] = -2\varphi' + 2\zeta \int_{r}^\infty dr' \varphi'^2(r')$
and the mobility takes the classical form
\begin{align}\label{eq:MAMB+}
    \mathcal{M}(R) = \frac{1}{4\pi (\Delta\phi)^2 R^3}.
\end{align} 
The critical radius is then
\begin{align}\label{eq:AMB+Rc}
    R_\mathrm{c} = \frac{2\sigma_B}{ \Delta g_B-f'(\phi_s)\Delta\psi }
\end{align}
where $\Delta g_B = g_B(\phi_s)-g(\phi_2)$ with $g_B(\phi)=\int^\phi d \varphi \, (\partial_{\varphi } f) \, e^{(\zeta-2\lambda)\varphi}$, and $\sigma_B=\int_r \varphi'[\psi'-\zeta(\psi(r)-\psi(\phi_2))\varphi']$.
The argument confirming that $\epsilon$ is small as assumed goes along similar lines as discussed above for the non-conserved case and in~\cite{prl} for the conserved one; we do not repeat it here. 

In summary, applying NNT to AMB+ gives Eq.~\eqref{eq:Rdot_intro} with the mobility and critical radius obeying Eqs.~(\ref{eq:MAMB+},\ref{eq:AMB+Rc}) respectively. The quasipotential is then found as
\begin{align}\label{AMB+UR}
U(R) = \frac{4\pi\sigma_B R^{2}\Delta \psi}{\Delta\phi}\left[ 1 - \frac{2R}{3 \, R_\mathrm{c}}\right].
\end{align}
Eqs.~\eqref{eq:MAMB+}--\eqref{AMB+UR} are the results originally obtained in~\cite{catesClassicalNucleationTheory2023}. 

It should be further noticed that, for AMB+ and other conservative models the same results are obtained by choosing $\psi'\notin\mathrm{Ker}(\mathcal{L}^\dagger)$ and simply setting $\epsilon=0$. (This is the TRR ansatz, made within the stochastic route.) Following the same derivation as before, we recover Eq.~\eqref{eq:Rdot_intro} with the mobility given by \eqref{eq:MAMB+}, the critical radius 
\begin{align}\label{eq:AMB+Rc-second}
R_\mathrm{c}=   \frac{2\bar\sigma}{ \Delta f_B-f'(\phi_s)\Delta\phi }\,,
\end{align}
and the quasipotential  
\begin{align}\label{eq:AMB+UR-second}
U(R) = 4\pi \bar\sigma R^{2}\left[ 1 - \frac{2R}{3 \, R_\mathrm{c}}\right]\,
\end{align}
where $\bar \sigma = \int_r \varphi'^2\left[1- \frac{3(2\lambda-\zeta)}{4} \varphi'_1 - \zeta(\varphi-\phi_2)\right]$.

Although not notationally obvious, the differences between
(\ref{eq:AMB+Rc-second}, \ref{eq:AMB+UR-second}) and (\ref{eq:AMB+Rc}, \ref{AMB+UR}) are subleading in the expansion in negative powers of $R$ and small supersaturation, so the two calculations are equivalent. This fact was already noticed in~\cite{catesClassicalNucleationTheory2023} so we do not prove it here.

\subsection{Action route for AMB+}\label{sub:AMB+TRR}
We finally show that the correct mobility, critical radius and quasi-potential, as respectively given by \eqref{eq:MAMB+}, \eqref{eq:AMB+Rc-second} and \eqref{eq:AMB+UR-second}, can also be obtained directly via the action route using the TRR ansatz which then reads $\epsilon_\mathrm{A}=0$. This confirms that, thanks to the conservation law, the difference between the density profiles on the instanton and the relaxational path does not affect the nucleation barrier to leading order.

The FW-action for AMB+ is given by
\begin{align}
    \mathcal{A} = -\frac{1}{4T}\int_{t,\p{x}} (\dot \phi - \nabla^2\mu)\nabla^{-2}(\dot \phi - \nabla^2\mu)\,.
\end{align}
We set $\epsilon_\mathrm{A}=0$ and thus set $\phi = \varphi(r-R)$. We keep the $\dot R$-dependent terms as, similarly to Eq.~\eqref{eq:action-TRS}, the $\dot R$-independent term in the time-reversed path vanish. Hence,
\begin{align}
    \mathcal{A} &= -\frac{1}{4T}\int_t \left( \int_\p{x} \varphi' \nabla^{-2}\varphi'\right)\times\nonumber\\ &\times\left[\dot R + \left( \int_\p{x} \varphi' \nabla^{-2}\varphi'\right)^{-1}\int_\p{x}\varphi'(\mu[\varphi]-f'(\phi_s))\right]^2
\end{align}
which is now the FW action for the $\dot R$ dynamics. Here, as before, we used the boundary condition on the Green function $\nabla^{-2}$ as $\mu(r\to \infty) = f'(\phi_\mathrm{s})$.
This implies that the mobility is
\begin{align}
    \mathcal{M}(R) = -\left( \int_\p{x} \varphi' \nabla^{-2}\varphi'\right)^{-1}\,.
\end{align}
Expanding now for large $R$ and large $R_\mathrm{c}$ and the interfacial shape as $\varphi=\varphi_0 + R^{-1}\varphi_1 + \mathcal{O}(R^{-2})$, we obtain Eq. \eqref{eq:Rdot_intro}, \eqref{eq:MAMB+}, \eqref{eq:AMB+Rc-second}, \eqref{eq:AMB+UR-second}. This shows that, for AMB+, we can obtain the nucleation barrier by considering the TRR path. This is always possible in equilibrium theories, but fails otherwise unless rescued by the conservation law, as happens here.

\section{Discussion}
\label{sec:discussion}
In this work, part of which was summarized in a companion article~\cite{prl},
we have extended Classical Nucleation Theory to nonequilibrium, nonconserved scalar field theories for a single order parameter that admits two homogeneous steady states. Our framework offers an analytical route to describe nucleation events in non-equilibrium systems, a problem that has previously been found generally intractable.

Three crucial features that allowed us to study nucleation in these non-equilibrium contexts are the following. First, we consider nucleation in a regime -- as classically done within CNT -- where the critical radius $R_\mathrm{c}$ is large. A perturbative structure is then present, that allows us to generalize CNT far from equilibrium, although the instanton trajectory is not the time-reversal of the relaxation path. Second, we considered situations where shape fluctuations of the nucleating droplet are resisted by a positive capillary-wave interfacial tension (that we computed in the explicit examples considered).  Since $R_\mathrm{c}$ is large, the resulting shape fluctuations are weak, justifying the spherical droplet assumption of NNT. 
Third, we employed a careful definition of the droplet radius $R$ (see Sec.~\ref{sec:int-position}). At first sight all reasonable definitions look equivalent since the resulting $R$ values differ by only of order unity whereas $R$ itself is large. However, only our chosen definition avoids the need to explicitly compute the deviation between the instanton trajectory and the time reversed relaxation path. 

We have derived NNT by two, mutually consistent approaches: (i) directly projecting the dynamics of the density field to obtain the Langevin equation for the droplet radius (stochastic route), and (ii) via minimization of the Freidlin–Wentzell action (action route).  We applied our general results to two examples drawn from active matter and population dynamics. Surprisingly, we find that deterministic nonequilibrium forces can enter the interfacial mobility. From the viewpoint of the stochastic route they do so via the function $\psi'$ used to optimally project the density field to the droplet radius. In the action route, they do so by causing a deviation in the radial density profile between that of a growing, instanton droplet and that of shrinking, relaxational one.  We have shown explicitly through the example of AMA that this deviation is large enough to substantively alter the quasipotential. Specifically it can reduce the barrier height for nucleation by  factors of at least 2 (see Fig.~\ref{fig:ama_together}).

To validate our theory, we performed direct numerical minimization of the action. 
We considered radially symmetric solutions to the instanton, which reduces the minimization to an effective one-dimensional problem, finding good agreement between the quasipotential barriers predicted from our NNT and the numerical minimization. The latter also directly detects deviations of the instanton from the time-reversed relaxation path that are consistent with our theory. 

The present framework is restricted to a single, non-conserved scalar field in the weak noise limit. 
Natural extensions include systems with multiple coupled order parameters; the inclusion of more complex boundary conditions~\cite{cavagna2024noise}; and mixed conserved and non-conserved dynamics \cite{Li_2020,Weber2019Apr,Zhou2026May}. 
Our work might thus be extended to describe nucleation for a broad class of problems, from ecology~\cite{Korniss2005Mar,DeAngelis2001Jan,Gandhi1999Sep,tanakaSpatialGeneDrives2017,giometto2021antagonism}, reaction diffusion systems~\cite{Ouazan-Reboul2023Jul,hellerPatternStabilityReactiondiffusion2026,cardy1998field,andreghetti2025enzyme}, and active systems with multiple (or vectorial/tensorial) order parameters. 
We expect NNT to serve as a general framework for nucleation in nonequilibrium systems, where analytical results have so far been scarce.

\begin{acknowledgements}
We thank Ronojoy Adhikari, Filippo De Luca, Robert Jack, and Nicolas Valade for fruitful discussions. We thank J. Tailleur for bringing~\cite{kuramoto1980instability,kawasaki1982kinetic,kawasaki1982kinetic-i,bausch1991effects} to our attention while this work was being finalized. CN acknowledges the
support of the ANR grant PSAM and the support of the INP-IRP grant IFAM. This research was supported in part by grant NSF PHY-2309135 to the Kavli Institute for Theoretical Physics (KITP).
We acknowledge the support and funding in part by the EPSRC through grant EP/Z534766/1.

\end{acknowledgements}

\bibliography{bibtex}
\end{document}